\documentclass[10pt,tightenlines,eqsecnum,floats,aps,amssymb,nofootinbib,prd,shownopacs,floatfix,twocolumn,superscriptaddress]{revtex4-2}

\usepackage[a4paper, total={7in, 9in},bottom=20mm,top=25mm]{geometry}
\usepackage{amsmath}
\usepackage{ulem}
\usepackage{tikz}
\usepackage{graphicx}
\usepackage{caption}
\usepackage{subcaption}
\usepackage{dsfont}
\usepackage{braket}
\usepackage[sort&compress]{natbib}
\usepackage{hyperref}
\usepackage{url}    
\usepackage{todonotes}
\makeatletter
\renewcommand{\p@subsection}{}
\renewcommand{\p@subsubsection}{}
\makeatother

\newcommand{\nocontentsline}[3]{}
\let\origcontentsline\addcontentsline
\newcommand\stoptoc{\let\addcontentsline\nocontentsline}
\newcommand\resumetoc{\let\addcontentsline\origcontentsline}

\begin{document}

\title{Horizon quantum geometries and decoherence}

\author{Max Joseph Fahn}
\email{maxjoseph.fahn@unibo.it}
\affiliation{Dipartimento di Fisica e Astronomia,
Universit\`a di Bologna, Via Irnerio 46,
40126 Bologna, Italy}
\affiliation{INFN Bologna, Via Irnerio 46, 40126 Bologna, Italy}

\author{Alessandro Pesci}
\email{pesci@bo.infn.it}
\affiliation{INFN Bologna, Via Irnerio 46, 40126 Bologna, Italy}

\begin{abstract}
There is mounting theoretical evidence that black hole
horizons induce decoherence on a quantum system, say a particle, put in a superposition of locations,
with the decoherence functional, evaluated after closure of the superposition, increasing linearly
with the time the superposition has been kept open.
This phenomenon has been shown to owe its existence
to soft modes, that is modes with very low frequencies, of the quantum fields --sourced by the 
particle-- which pierce through the horizon,
or also can be understood as coming from the interaction 
with the black hole described as a thermodynamic quantum system at Hawking's temperature.

Here we investigate the effects of ensuing quantum 
aspects of the geometry itself of the horizon,
in an effective perspective in which the quantum 
geometry of the horizon is captured by existence of a limit length or by horizon area quantisation.
We show that the discreteness of the energy levels 
associated to the different geometric configurations
might have strong impact on the results,
in particular reducing the decoherence effects even
to a negligible level in case of quanta of area $A_0 = \mathcal{O}(1) \, \, l_p^2$ or larger,
with $l_p$ the Planck length.

\end{abstract}

\maketitle

\section{Introduction}

Recent work has shown that black hole horizons do induce decoherence on a system in quantum superposition of locations \cite{Danielson:2022tdw, Danielson:2022sga, Gralla:2023oya, Wilson-Gerow:2024ljx, Danielson:2024yru},
this having been epitomized by the phrase: ``Black Holes are Watching You'' \cite{DICEWald}.
%\textcolor{blue}{(this corresponding to the general expectation
%that curvature affects the coherence of quantum probes,
%see e.g. \cite{singh2023decoherence} for a recent account).} 

This result has its roots in previous findings coming from consideration of a, well-known by now, gedankenexperiment,
in which 
a delocalized mass $A$ is recombined while another mass $B$ probes its gravitational field,
or, analogously, 
a delocalized charge $A$ is recombined while another charge $B$ probes its electromagnetic field
(put forward in \cite{Mari:2015qva, Belenchia:2018szb, WalG, WaldH}, on a previous proposal in \cite{Baym:2009zu}).
In this setup a possible tension between causality
and complementarity (the latter being the fact that knowing the path means
losing the interference pattern at recombination) arises,
if the mediating field is quantum, 
when the two masses or charges happen to be causally disconnected.
Indeed, if $B$, thanks to the fact that the mediating field is quantum,  
can discriminate the path
while $A$ recombines, and all this
happens in a time short enough to be smaller than the light time between $A$ and $B$, how can $A$ come to know of B's success in path discrimination? 

Sticking to the gravitational case,
the solution proposed in \cite{Belenchia:2018szb}
(other accounts, with different perspectives, are in
\cite{Rydving, Grossardt, Pesci:2022vwm})
is brilliant: 
When conditions are such that
(read: when mass $m$ and separation $d$ of $A$ are large enough)
$B$ can give which path (assuming ideal 1 Planck length $l_p$ resolution) in a time smaller than the light time $t_l$ between the two particles,
then if, for that same conditions, $A$ also recombines in a time $< t_l$ (in order not to know of the successful
path discrimination by $B$),
then
$A$ necessarily emits a graviton and thus decoheres.
This aligns with the solution proposed in \cite{Baym:2009zu} for the electromagnetic case,
where charge replaces mass and one has emission of a photon instead of a graviton. 
No need of information then, going from
$B$ to $A$, to get $A$ decohering;
the graviton is emitted no matter what $B$ does.

What has been done in \cite{Danielson:2022tdw, Danielson:2022sga} is sort of considering a situation where $B$ is located behind a horizon. In that case $B$ can indeed discriminate the path as before
(thanks to the information which comes through the horizon from outside), 
but $A$ can now recombine in a very long time
still knowing nothing from $B$ for no signal
can come to $A$ from $B$,
and this challenges
the resolution of the paradox as along the lines described above.

The proposed resolution is that the mere presence of the horizon needs to induce decoherence in the superposition of $A$.
This was shown in \cite{Danielson:2022tdw, Danielson:2022sga, Gralla:2023oya} considering  the global state of the
quantum field to which both $A$ and $B$ coupled.
The key result has been that the soft modes of the field
piercing a Killing horizon lead to a loss of coherence
in $A$ superposition. 
In \cite{Wilson-Gerow:2024ljx, Danielson:2024yru} 
the same problem
was faced, now from the local point of view of $A$.
In this approach, the presence of the horizon
manifests in a modified form of the fluctuation spectrum of the quantum field in $A$'s lab, inducing the
decoherence.
All these works have in common that
they treat spacetime as classical, in particular they
assume that any modes of the field, independent of the frequency associated to them, 
can pierce and enter, or leave 
(for the quantum vacua), the horizon. 
Aim of the present investigation
is to try to reconsider those results
for the case of the implementation of specific quantum features 
of the horizon.
It is intended as a detailed
description and further development of the ideas sketched in \cite{FahnP}.

Due to the fundamental result by Hawking \cite{Hawking},
black holes do behave as thermodynamic objects
and as such they have to be regarded as fundamentally
quantum.
In this sense the results just mentioned do refer
already to quantum black holes, 
in spite of spacetime being treated as classical,
and this is 
also the perspective advocated by \cite{Biggs:2024dgp}
(and, specifically for extremal Reissner-Nordstr{\"o}m black holes, in \cite{Li:2025vcm})
in which the black
hole is actually regarded as a thermal black quantum system which induces decoherence on the system in superposition it is coupled to, much like 
it would an ordinary piece of matter at finite temperature, and the same results \cite{Danielson:2022tdw, Danielson:2022sga, Gralla:2023oya, Wilson-Gerow:2024ljx, Danielson:2024yru}  
are re-obtained. These results were also confirmed in the context of holography, see \cite{Kawamoto:2025kfu}.

Starting from this,
what we would like to 
do here is to ask a further question.
If the black hole is a thermodynamic quantum system, it has clearly, entropy, and discretized energy levels, corresponding to quantum aspects of the spacetime.
Our aim is to explore the consequences of a quantum nature
of the spacetime itself associated to the black hole.  

The consideration of the quantumness of spacetime might bring in no surprises, in the sense that those same results one gets considering the black hole as generically a quantum system at finite temperature
can be re-obtained, exactly the same, even after explicit consideration
of the discrete nature of the allowed energy levels.
It might be however that certain results do depend
in a crucial manner on the possibility to consider exchange of energies which are vanishingly small,
which is actually what seems to happen in the results above in which the linear increase of decoherence arises
decisively from the soft modes absorbed by the horizon.
In this second case, the discretization of energy levels,
in particular the existence of an elemental quantum of energy below which we can not have energy exchange, might have impact on the results. This is actually what we would like to investigate. 

For quantum black holes (quantum black hole spacetimes) 
indeed, a general argument
going back to long ago foresees
a discrete energy spectrum with quantization
of their area, with a behavior akin to 
atomic systems \cite{Bekenstein}.
Black hole area quantization comes also
from full-fledged quantum theories of
gravity, like e.g. from loop quantum gravity 
\cite{Ashtekar}.
The prediction of the value of the quantum of area is somewhat more ambiguous, going from $8\pi l_p^2$
for the Bekenstein's model with equally spaced levels
(other accounts, from ringdown modes, are in \cite{Hod, Maggiore}) 
to non equally spaced levels in loop quantum gravity with gaps possibly decreasing
in size with the area of the black hole giving rise
to an almost continuum area spectrum for reasonable
black hole areas \cite{Barreira, Barbero} (there is however
a calculation in loop quantum gravity which
foresees equally spaced levels with gap 
$(4 \ln 3) l_p^2$ \cite{Barbero2}).

In the following we will describe quantum horizons 
in an effective manner using the minimum-length metric, namely a metric description of spacetime with a minimum-length $L$ built-in \cite{KotE, KotF, KotI}.
This effective metric predicts indeed a gap in area of calculable amount, growing with the parameter $L$ of the metric and vanishing when $L=0$,
and which turns out to be of the same order of magnitude of Bekenstein's prediction above when one takes $L = l_p$ \cite{KriPer, DICE}.

The structure of the paper is as follows.
In Section~\ref{sec:ClasHor} we consider classical horizons and recall
some of the techniques used in the works \cite{Danielson:2022tdw, Danielson:2022sga, Gralla:2023oya, Wilson-Gerow:2024ljx, Danielson:2024yru},
both in the global and the local approach, 
with some elaboration on them; we shall put particular emphasis on those aspects of the calculations 
which, as a matter of fact, will be affected once the horizon area is considered as quantized.
In Section \ref{sec:quantum} we investigate then the effects on the decoherence functional arising from having the horizons quantized.
It follows then a discussion of the results,
and the conclusions.
We work in mostly positive signature, and use 
units with $G=\hbar=c=1$
unless, at times, explicitly stated otherwise.

\section{Decoherence induced by horizons with underlying classical 
geometry}\label{sec:ClasHor}

The works that discuss the decoherence induced by a classical horizon 
address this effect from different perspectives.
The discussions in \cite{Danielson:2022tdw,Danielson:2022sga,Gralla:2023oya} follow a global point of view: they determine the decoherence by calculating the number of particles radiated from $A$'s superposition through the horizon. In \cite{Danielson:2022tdw}, 
for a (Schwarzschild) black hole, 
first focusing on the electromagnetic case the electric field is split at asymptotically late times into a Coulomb part, a radiation part (which is assumed to produce negligible decoherence) and the electromagnetic field state on the event horizon, which is expected to yield the dominating contribution to the decoherence of $A$'s superposition. From the Maxwell equations on the horizon follows then the radiation that pierces the horizon, which allows to determine the decoherence using Fock quantization on Schwarzschild background and assuming the electric field to be in Hartle-Hawking vacuum (which behaves similarly to the Unruh vacuum for low frequencies near the horizon). The ``soft photons", i.e. the very low frequency radiation, then lead to decoherence that is proportional to the time the superposition was kept open (assuming the opening and closing happens much faster, but still adiabatically):
\begin{equation}
    \langle N \rangle \propto \frac{M^3 q^2 d^2}{D^6} {\cal T}\,.
\end{equation}
Here, $\langle N \rangle$ denotes the number of radiated particles through the horizon (including also the soft photons), where a value of $\langle N \rangle \gtrsim 1$ leads already to considerable decoherence. Furthermore, $M$ is the mass of the black hole, $q$ is the charge and $d$ the spatial separation of $A$'s superposition, and $D$ denotes the proper distance of $A$'s lab to the horizon. 
$\cal T$ is the time the superposition has been
kept open.
Due to close analogy, in \cite{Danielson:2022tdw} is discussed that these results directly carry over to the case when replacing the electric field with a linearized gravitational field.

In \cite{Danielson:2022sga}, the results from \cite{Danielson:2022tdw} are generalized to spacetimes with Killing horizons, in particular to stationary black holes, Rindler horizons and cosmological horizons in a de Sitter spacetime. A similar procedure as in \cite{Danielson:2022tdw} leads to decoherence that grows linearly with the time $A$'s superposition was kept open if $A$ follows a similar protocol opening and closing her superposition. 
In the Rindler case,
in particular,
it is found
\begin{equation}
    \langle N \rangle \propto \, a^3 q^2 d^2 {\cal T}\,,
\end{equation}
for the electromagnetic field,
with $a$ the acceleration of $A$'s lab.

The precise version of the 
order-of-magnitude scaling
in the formulas above
has been provided in
\cite{Gralla:2023oya}
for the electromagnetic
case
(see Eq. \eqref{eq:Dfctapro1} below)
and extended to include the case of
a scalar field.
\cite{Gralla:2023oya} also explicitly calculates
the decoherence rate $A$ observes if placed on the symmetry axis of a Kerr black hole.
The result intriguingly shows 
in particular that in the
extremal limit 
no decoherence is induced by the
horizon, this in turn corresponding to the
fact that in this limit
no external fields can enter the horizon.

The subsequent works \cite{Wilson-Gerow:2024ljx,Danielson:2024yru} take a local point of view and analyze the decoherence by evaluating the quantum field in $A$'s lab. These works show in particular that the global and local approaches lead to equivalent physical predictions. Starting with $A$'s lab following a Rindler trajectory in Minkowski spacetime, \cite{Wilson-Gerow:2024ljx} shows that the superposition experiment can be equally described by $A$ opening and closing a dipole and thereby connects it to an Unruh-DeWitt detector. The influence functional approach \cite{Feynman:1963fq} allows then to describe this dipole as an open quantum system, treating the electric (or linearized gravitational) field as environment. The decoherence is then obtained by evaluating the two-point correlation functions of the electric field evaluated on $A$'s Rindler trajectory. Again, the contribution relevant for the decoherence effect, linear in the time the dipole was kept open, comes from the very low frequency modes of the electric field, which confirms the results found from the works following the global perspective. In particular, similar to the outcome in \cite{Danielson:2022sga}, 
scattering of Unruh radiation is not sufficient to lead to this decoherence effect.

In \cite{Danielson:2024yru} this local point of view is enhanced and applied to the black hole spacetime,
where the two-point correlation functions of the electric field (quantized here in Schwarzschild) are now approximately evaluated in $A$'s lab, 
and the differences of Hartle-Hawking, Unruh and Boulware vacua are also discussed. 
While the former two yield the same result for decoherence as described above (due to their similar behavior at low frequencies close to the horizon), the latter leads to only a logarithmic dependence of the decoherence on the (large) time the superposition was kept open.

In \cite{Biggs:2024dgp}, $A$'s dipole couples to a ``black quantum system" 
(modeling the quantum
black hole)
and the system is described by an effective theory obtained from integrating out the electric/linearized gravitational field's degrees of freedom, thereby yielding a dipole-dipole (electric case) or higher-order multipole interaction. Similar to the local point of view, the decoherence can then be determined by an evaluation of the two-point correlation functions of the black quantum system's multipole operators. Matching Schwarzschild calculations with the effective black quantum system model allows to determine these functions for a black hole and reproduces the decoherence rates obtained in the previous works. In the end, the question whether ordinary matter can reproduce a similar decoherence effect is analyzed and affirmed for the electric, but rather difficult to obtain for the linearized gravitational case. This black quantum system description is also pursued in \cite{Li:2025vcm}, where near-extremal Reissner-Nordström black holes are discussed and it is shown that the decoherence can be partially enhanced at low temperatures if one describes the correlation function as arising from a canonical ensemble.

In what follows, to better appreciate
what will happen later when introducing a quantum geometry,
we first briefly recap the idea of the global approach in section \ref{sec:globap}, and describe the local approach
(which will be our workhorse when considering the quantum geometry)
in section \ref{sec:locap}.  
Our focus is on the electromagnetic case.

\subsection{Estimate of the decoherence: global approach}\label{sec:globap}

In this section we recall a possible derivation of the decoherence $A$'s superposition exhibits due to the presence of a horizon from the global point of view. The presentation follows closely \cite{Danielson:2022sga,Gralla:2023oya} and summarizes some of their results.

At a point $P$ in a stationary black hole spacetime 
(and assumed asymptotically Minkowski) which we take here
Schwarzschild,
we consider a charged particle
which is put
in a quantum superposition of locations $\ket{\psi_L}$ and $\ket{\psi_R}$,
then described by the state
\begin{eqnarray}
    |\psi\rangle = 
    \frac{1}{\sqrt{2}} \big(|\psi_{L}\rangle + |\psi_{R}\rangle\big)
\end{eqnarray}
and, after a (Schwarzschild) time interval $\Delta t \equiv T$,  
goes to recombine.
We assume that the superposed states are orthogonal to each other and that the electromagnetic current $j^\mu$ associated to this charge is essentially classical (including the currents $j^\mu_L$, $j^\mu_R$ associated to the two branches)   
i.e., they have quantum fluctuations negligible against the expected values
and do not exhibit appreciable reaction from radiation.
This is assumed to be the case both in the stationary phases (before separation, in the time $T$ in which the separation is maintained, after recombination) 
and in the transient phases of separation and recombination.

The evolution we imagine for the system is such that at early and late times we have $j^\mu_L = j^\mu_R$.
The quantum electromagnetic field sourced by
these currents is in the form
of coherent states $|A^\mu\rangle$ \cite{Glauber}.
At start the particle is not yet delocalized and the particle+field system is described by the (Schr\"odinger picture) state
\begin{eqnarray}
|\chi\rangle = |\psi_i\rangle |A_i\rangle,
\end{eqnarray}
with $|\psi_i\rangle$, $|A_i\rangle$ the particle and its sourced field states (dropping spacetime indices in the field states for ease of notation).
With the delocalization just happened, say at $t = t_s$, let the particle+field system be in the state
\begin{eqnarray}
    |\chi\rangle = \frac{1}{\sqrt{2}} \big(|\psi_L\rangle + |\psi_R\rangle\big) \, |A_i\rangle.
\end{eqnarray}
This corresponds to the assumption that at the end of the separation process the fields of the two branches, $|A_{L, R}\rangle$ 
associated to the currents $j_{L,R}$ are still
$\approx |A_i\rangle$, that is the experimenter has been able to perform the separation almost coherently, 
without appreciable influence by the environment. This may or may not be satisfied in actual circumstances, but the point of the paper is in principle to see what happens when
the transients in the opening and closing of the superposition may have negligible effects; we will 
discuss this in some more detail later on, in particular the limits of applicability in our framework. 
During delocalization, the state then evolves as
\begin{eqnarray}
    |\chi\rangle = \frac{1}{\sqrt{2}} 
    \big(|\psi_L\rangle \, |A_L\rangle
    + |\psi_R\rangle \, |A_R\rangle\big).     
\end{eqnarray}

This state can be usefully expressed in terms of the density matrix $\rho_{p+f}$ of the particle+field system as
\begin{eqnarray}
  \rho_{p+f} = 
  \frac{1}{2} \Big(|\psi_L\rangle |A_L\rangle \langle\psi_L|\langle A_L| +
  |\psi_L\rangle |A_L\rangle \langle\psi_R|\langle A_R|
  \nonumber \\
  +|\psi_R\rangle |A_R\rangle \langle\psi_L|\langle A_L| +
  |\psi_R\rangle |A_R\rangle \langle\psi_R|\langle A_R|\Big).  \nonumber
  \end{eqnarray}
The reduced density matrix for the particle alone, obtained by tracing out the field, then reads
\begin{eqnarray}\label{rho}
\rho_p &=& \frac{1}{2} \Big( |\psi_L\rangle\langle\psi_L| +\langle A_R|A_L\rangle |\psi_L\rangle\langle\psi_R| \nonumber \\
&+& 
\langle A_L|A_R\rangle |\psi_R\rangle\langle\psi_L|
+ |\psi_R\rangle\langle\psi_R|\Big).
\end{eqnarray}
Decoherence $\cal D$ corresponds to the destruction of the off-diagonal elements and can be defined as 
\begin{equation}\label{deco}
{\cal D} = 1 - |\langle A_L| A_R\rangle|.
\end{equation}
At start i.e., when separation has just begun, these have almost their maximum value $|\langle A_L| A_R\rangle \approx |\langle A_i| A_i\rangle| = 1$, and what one is interested in is $|\langle A_L | A_R\rangle |$ when the particle is recombined.

This quantity can not change anymore after the recombination has been performed, and
works \cite{Danielson:2022tdw,Danielson:2022sga,Gralla:2023oya}
have shown 
(actually, for a Kerr generic black hole, with $t$ the coordinate time)
that it can be conveniently computed in the $t\to\infty$ limit, as
\begin{eqnarray}\label{exp-N/2}
|\langle A_L | A_R\rangle | \, = \, e^{-\langle N\rangle/2},
\end{eqnarray}
with
\begin{eqnarray}\label{N}
    \langle N \rangle = 
    \lim_{t\to\infty} \big(K \Delta A, K \Delta A\big)_t.
\end{eqnarray}
Here
$\Delta A = \big(A_R^{ret} - A_R^{adv}\big) - \big(A_L^{ret}-A_L^{adv}\big)$, with
$A_{L,R}^{ret}$ the retarded and $A_{L,R}^{adv}$ the advanced solutions to $\nabla_\mu F^{\mu\nu} = - 4\pi j^\nu_{L,R}$ (with $F_{\mu\nu} = \partial_\mu A_\nu - \partial_\nu A_{\mu}$), 
and $K$ means taking the positive frequency part.
The expression under the limit on the R.H.S. of
(\ref{N}) is the Klein-Gordon-like product
of solutions to field equations at time $t$, namely
\begin{eqnarray}
    (A_1, A_2)_t = 
    -\frac{i}{4\pi} \int_{\Sigma_t} d^3 x \sqrt{h} \, n^\mu \big({\bar F}_{1 \mu\nu} A_2^\nu - F_{2 \mu\nu} {\bar A}_1^\nu\big) \nonumber
\end{eqnarray}
with $n^\mu$ the normal to a spacelike Cauchy surface
$\Sigma_t$, $\sqrt{h} d^3 x$ the volume element in terms of the induced metric $h_{ij}$, this product being independent on $t$ for  solutions of the homogeneous field equations, as the retarded minus advanced solutions are.
As the notation suggests $\langle N \rangle$ is a number of particles; it turns out to equal the expected number of particles
in the out state (i.e., after recombination) corresponding to the in vacuum (i.e., to before separation), 
and we see they are sourced by $j_R - j_L$. 

For fields which drop out quickly enough in the $t\to~ \infty$ limit, which, as it turns out, is appropriate for our system \cite{Danielson:2022tdw,Danielson:2022sga,Gralla:2023oya},
the limit product in (\ref{N}) 
is expressible in terms of two contributions, one from the horizon $H$ and the other from null infinity $\cal I$
\begin{eqnarray}
\lim_{t\to\infty} \big(K \Delta A, K \Delta A\big)_t &=&  \nonumber \\
\big(K \Delta A, K \Delta A\big)_H 
&+& 
  \big(K \Delta A, K \Delta A\big)_{\cal I}\:,      
\end{eqnarray}
the latter moreover turning out to be 
negligible 
for our electromagnetic dipole field $j_R - j_L$ if choosing appropriate transients \cite{Danielson:2022tdw,Danielson:2022sga,Gralla:2023oya},
see also \cite{Belenchia:2018szb}.
The horizon contribution can be computed at future, or at past horizon. Taking positive frequencies with respect to horizon-generators affine time corresponds for the past horizon to assume the Unruh state \cite{Unruh}
which means to have boundary conditions adequate for the black hole having formed from gravitational collapse.   

The end result can be conveniently expressed in the form \cite{Gralla:2023oya}
\begin{eqnarray}\label{N_2}
    \langle N\rangle &=&
    \frac{1}{4 \pi^2} \int_{\cal B} dS \int_0^{\bar\omega} \frac{\omega d\omega}{2\pi} |\Delta{\widetilde A}|^2 \coth\Big(\frac{\pi \omega}{\kappa}\Big)
    \nonumber \\
    &=&
    \frac{2}{\pi} C \int_0^{\bar\omega} \frac{d\omega}{\omega} \sin^2\Big(\frac{\omega T}{2}\Big) \coth\Big(\frac{\pi \omega}{\kappa}\Big)
\end{eqnarray}
with large $T$ limit implied,
\begin{eqnarray}
    C = \frac{1}{4 \pi^2} \int_{\cal B} dS \ {\widehat {\Delta A}}^2,
\end{eqnarray}
and $\bar\omega$ a cut frequency definitely $\bar\omega \gg \frac{1}{T}$
(which can be thought as effectively expressing the effects of a smooth opening and closing of the dipole) and $\bar\omega \lesssim \kappa$,
with the result of the integration being independent of $\bar\omega$. 
Here the $dS$ integration is on the 2-dim spacelike bifurcation surface $\cal B$; $\kappa$ is horizon surface gravity and $\omega$ is the angular frequency associated to Killing time, namely to the horizon generating Killing field $\xi^\mu$. This, for Schwarzschild, namely the case we focus on, is $\xi^\mu = \big(\frac{\partial}{\partial t}\big)^\mu$, i.e., the Killing time is just $t$.

$\widehat {\Delta A} = \Delta A^{stat}_{|H}$ 
is the constant vector
associated to the stationary solutions $A_{L, R}^{stat}$ 
in the horizon-adapted gauge \cite{Danielson:2022tdw,Danielson:2022sga,Gralla:2023oya}
(i.e., with $A^\mu$ vanishing when contracted with the horizon generators) sourced by the stationary currents $j^{stat}_{L,R}$ all along the superposition time.
$\Delta \widetilde A$ in the first equality of (\ref{N_2}), 
is the Fourier transform of $\Delta A_{|H}$ with respect to Killing time
assuming a top hat evolution for $\Delta A_{|H}$ with height $\widehat {\Delta A}$, namely, for Schwarzschild,
\begin{eqnarray}
    \Delta\widetilde A &=& \int_{-\infty}^{+\infty} \Delta A_{|H} \ e^{-i \omega t} dt
    \nonumber \\
    &=& \widehat {\Delta A} \ \frac{2}{\omega} \ \sin\Big(\frac{\omega T}{2}\Big).
\end{eqnarray}

The final result (\ref{N_2}) turns out to be essentially the same one finds if the description of the process is done in terms of the quantities directly accessible by the experimenter at $P$, that is according to a local description as opposed to a global one.
This corresponds to something one would expect: Given a stationary spacetime endowing a horizon, the effects of the horizon on the evolution of a quantum system (the quantum charge) held stationary which are captured considering the global evolution of the sourced fields,
can also be captured by consideration of the 2-point function of the quantum field where the superposition is,
\cite{Wilson-Gerow:2024ljx, Danielson:2024yru}.
In \cite{Danielson:2024yru}, this local description is applied in a Schwarzschild spacetime by an approximate evaluation of the 2-point functions in Boulware \cite{Boulware:1974dm}, Unruh \cite{Unruh} and Hartle-Hawking \cite{Hartle:1976tp} vacuum states.
In \cite{Wilson-Gerow:2024ljx}, the local description is done for a Rindler spacetime. Our aim here is to use the latter in the 
near-horizon limit, namely as the local Rindler frame which approximates the black hole horizon, i.e., where
the decoherence effects by the horizon might be expected to
be the strongest.
In the next section we briefly recall how the local description 
comes about following 
specifically
the derivation of~\cite{Wilson-Gerow:2024ljx}.

\subsection{Estimate of the decoherence: local approach}\label{sec:locap} 

The idea in \cite{Wilson-Gerow:2024ljx,Danielson:2024yru} is to use the theory of open quantum systems (see e.g. \cite{breuer2002theory} for an introduction) for an analysis of the situation in \cite{Danielson:2022tdw,Danielson:2022sga,Gralla:2023oya,Danielson:2024yru} by considering $A$'s particle as accelerating \cite{Wilson-Gerow:2024ljx}
and modeling the particle's spatial superposition in terms of an accelerating electric dipole which is created during some  proper time ${\cal T}_1$, then kept open for an interval ${\cal T} \gg {\cal T}_1$ and closed afterwards during a time interval which we also choose to be ${\cal T}_1$, all this in the Rindler frame of the accelerating particle.
As mentioned,
in the context of a black hole
this frame should be regarded 
as the approximating local
Rindler frame at particle location
of the actual spacetime 
endowing the stationary horizon.
In \cite{Wilson-Gerow:2024ljx} it is shown that this setup is sufficient to restore the physical case considered in \cite{Danielson:2022tdw,Danielson:2022sga}.
In what follows, we briefly summarize some of the main steps from \cite{Wilson-Gerow:2024ljx} in order to set the stage for the implementation of a quantum horizon geometry in section \ref{sec:quantum}.

\subsubsection{The decoherence functional}
 Following \cite{Wilson-Gerow:2024ljx}, 
 we consider a lab accelerating with the particle 
 and describe the evolution of the system in terms of the
 associated (local) Rindler frame with acceleration $a$.
 Particle $A$ is
 modeled 
 as a two-state system
 with separation $d\ll \frac{1}{a} = D$, with $D$ the distance
 to the horizon,
 where the two states $L$ and $R$ correspond to the trajectory of either end of the dipole, 
 and can be described in the following terms
\begin{equation}\label{rho_2}
    \hat{\rho}_A= \begin{pmatrix}
        \rho_{LL} & \rho_{LR} \;\mathcal{F}[L,R] \\
        \rho_{RL} \;\mathcal{F}[R,L] & \rho_{RR}
    \end{pmatrix}\,,
\end{equation}
where $\rho_{LL}$ and $\rho_{RR}$ are the populations corresponding to states $L$ and $R$, that is the probabilities that the particle is found in one of the two branches if measured. The off-diagonal elements denote interference probabilities and $\mathcal{F}$ is the Feynman-Vernon influence functional \cite{Feynman:1963fq}. Its absolute value is defined as 
\begin{equation}\label{deco_2}
    |\mathcal{F}| = e^{-\frac{1}{2}\mathds{D}}
\end{equation}
($|\mathcal{F}[L,R]| = |\mathcal{F}[R,L]| \equiv |\mathcal{F}|$) 
with the decoherence functional $\mathds{D}\geq 0$. It can be readily seen that the latter, if non-zero, damps the interference probabilities and thus leads to decoherence of $A$'s particle. The decoherence functional encodes the influence of the environment, in the present case of the quantum field (which sees the horizon and thus contains the information of its presence).

\subsubsection{Decoherence in a first approximation}\label{sec:decfa}

Similarly to \cite{Wilson-Gerow:2024ljx}, we denote with $X^\mu$
the position of the lab (center of mass of the two branches) in the Minkowski frame (taking the $Z$-axis along the acceleration direction) 
\begin{equation}\label{eq:traj}
    X^\mu(\tau) = \frac{1}{a} \left(\sinh(a \tau), 0, 0, \cosh(a\tau) \right)\,,
\end{equation}
with $a$ denoting the proper acceleration and $\tau$ the proper time,
and the dipole moment
\begin{equation}\label{eq:dipmomsolx}
    q \epsilon^\mu(\tau) =q (\epsilon_3(\tau) \sinh(a\tau),\epsilon_1(\tau),\epsilon_2(\tau),\epsilon_3(\tau) \cosh(a \tau)),
\end{equation}
with $\epsilon_A$, $A = 1,2,3$, the components of the separation vector in the orthonormal lab frame
(with the 3-axis along $Z$).
With this, the decoherence functional
reads
\begin{align}\label{eq:defdecfctft}
    \mathds{D} &= \frac{q^2}{2} \int_{-\infty}^\infty d\tau\, d\tau'\; \epsilon^A(\tau)\; \epsilon^B(\tau') \langle\{E_A(X(\tau)),E_B(X(\tau'))\}\rangle \nonumber\\ &=\frac{q^2}{2} \int_{-\infty}^\infty \frac{d\Omega}{2\pi}\; S_{AB}(\Omega) \; \tilde{\epsilon}^A(-\Omega) \, \tilde{\epsilon}^B(\Omega)\,
\end{align}
(omitting spacetime indices for $X^\mu$),
where $E^A(\tau)$
denotes the electric field in the 
lab frame,
$S_{AB}$ is the Fourier transform of the two-point correlation function of the electric field (for which one can use that the correlation function here depends only on $\tau - \tau'$, as the environment initially is in the Minkowski vacuum state which is a thermal state, see \cite{Wilson-Gerow:2024ljx}) and $\tilde{\epsilon}^A$ denotes the Fourier transform of $\epsilon^A$, $\tilde{\epsilon}^A(\Omega) = \int_{\mathds{R}} d\tau e^{-i\Omega\tau} \epsilon^A(\tau)$.  
The Fourier transform of the two-point correlation function is 
with respect to $\Delta\tau \equiv \tau - \tau'$
\cite{Wilson-Gerow:2024ljx}
\begin{align}\label{eq:SAB}
    S_{AB}(\Omega) &= \int_{-\infty}^{\infty} d(\Delta\tau)\,   e^{-i\Omega(\Delta\tau)}  \langle\{E_A(r(\tau)),E_B(r(\tau'))\}\rangle\ \nonumber\\ &= \delta_{AB} \frac{\Omega^3 + a^2\Omega}{3\pi \epsilon_0} \; \coth\left(\frac{\pi\Omega}{a}\right)\,.
\end{align}
The assumptions that enter in this derivation 
of $\mathds{D}$
are that higher order correlation functions can be neglected and that it is sufficient to evaluate the influence functional along the classical trajectory of Alice's particle.

One of the key results in \cite{Wilson-Gerow:2024ljx} is that this correlation function deviates from the Planck spectrum for thermal fluctuations, where the term linear in $\Omega$ is absent. This linear term is the crucial ingredient for the different decoherence obtained in the presence of the horizon in comparison to a similar dipole in a thermal environment, this horizon being meant as the Rindler horizon approximating the actual black hole horizon.

To evaluate the decoherence functional \eqref{eq:defdecfctft}, it is now only required to give some more details on the dipole moment $\epsilon_A(\tau)$. 
For a dipole kept open, as contemplated above,
for a very long time ${\cal T}$, while its creation and destruction time ${\cal T}_1 \ll {\cal T}$ is taken to be relatively small, 
the Fourier transform $\tilde{\epsilon}_A(\Omega)$ is strongly band-limited with a narrow bandwidth around $\Omega=0$, $\tilde{\epsilon}_A(\Omega) = 2d\, \frac{\sin\left( \frac{\Omega {\cal T}}{2}\right)}{\Omega} \hat v_A$  
(with $\hat v_A$ the unit (spatial) vector in the direction of the separation vector),
analogously to what was considered above.
Using this, and assuming that $\tilde\epsilon$ has only frequencies
$\Omega < \bar\Omega$ (from smoothness in the opening and closing of the dipole) 
one gets (cf. \cite{Wilson-Gerow:2024ljx})
\begin{equation}\label{doubleD}
    \mathds{D} = \frac{2}{\pi a^2}  \ C \int_0^{\bar\Omega} d\Omega  \ \frac{a^2 + \Omega^2}{\Omega} \sin^2\left(\frac{\Omega {\cal T}}{2}\right) \coth\left(\frac{\pi \Omega}{a}\right)\
\end{equation}
with $C$, as it turns out, the same constant as in equation 
(\ref{N_2}) above. 
From this, when $\bar\Omega \ll a$
one obtains 
\begin{equation}\label{doubleD_2}
     \mathds{D} = 
     \frac{2}{\pi}  \ C \int_0^{\bar\Omega} d\Omega  \ \frac{1}{\Omega} \ \sin^2\left(\frac{\Omega {\cal T}}{2}\right) \coth\left(\frac{\pi \Omega}{a}\right)\,,
\end{equation}
the result of the integration turning out to be independent of $\bar\Omega$
in the large $\cal T$ limit.
In spite of the different frequencies and times,
this expression is actually equal to (\ref{N_2})
in the near-horizon limit we are in,
as can be readily seen by the fact that in this limit
$\kappa \, dt = a \, d\tau$ and then 
$\omega = \frac{\kappa}{a} \, \Omega$
(cf. \cite{Gralla:2023oya}).
Looking at (\ref{deco_2}, (\ref{rho_2}), and at 
(\ref{exp-N/2}), (\ref{rho}) we see that $\mathds{D}$ has to play the role of $\langle N \rangle$.
It has indeed the meaning of number of emitted particles by the dipole in the local description \cite{Wilson-Gerow:2024ljx}.

The factor 
$\frac{a^2+\Omega^2}{a^2 \Omega} =
\frac{1 \, + \, \Omega^2/a^2}{\Omega}$ in \eqref{doubleD} in contrast to $\frac{1}{\Omega}$ 
we would have from \eqref{N_2} can be understood as due to different approximations invoked in \cite{Wilson-Gerow:2024ljx} and \cite{Gralla:2023oya}. 
While the integrands agree in the limit of very small $\Omega$ or $\omega$, which is the main subject of discussion in these works for horizons with classical geometries, 
in the present work, 
which precisely challenges this limit,
we choose to work with expression \eqref{doubleD} from \cite{Wilson-Gerow:2024ljx}. 
Since for the discussion of horizons with quantum geometries 
it will be sufficient to focus on $\bar\Omega \sim a$, 
we see this only amounts
to some (unessential, in a sense which will become clear in the following)
$\mathcal{O}(1)$ factor 
between the two expressions.

In the small-$\Omega$ limit equation
\eqref{eq:defdecfctft} shows that
\cite{Wilson-Gerow:2024ljx}
\begin{align}\label{eq:Dfctapro1}
    \mathds{D} 
    &\approx \frac{q^2}{2} S_{AB}(0) \int_{\mathds{R}} \frac{d\Omega}{2\pi} \; \tilde{\epsilon}^A(-\Omega) \tilde{\epsilon}^B(\Omega)\nonumber\\ &= \frac{q^2}{2} S_{AB}(0) \int_{\mathds{R}} d\tau \; {\epsilon}^A(\tau) {\epsilon}^B(\tau) \nonumber\\ &\approx \frac{q^2}{2} \epsilon^A \epsilon^B \, {\cal T} \, S_{AB}(0) = \frac{q^2d^2 a^3}{6 \pi^2
    \epsilon_0} \, {\cal T} 
    = \frac{a}{2\pi} \, C \, {\cal T}\,,
\end{align}
where the last equality is from (\ref{doubleD_2}) in the large $\cal T$ limit.
This yields a decoherence functional scaling linearly with the time $\cal T$ the superposition has been kept open. 
This confirms the results from \cite{Danielson:2022tdw,Danielson:2022sga,Danielson:2024yru} (where the proportionality was determined without prefactors) and, as we have seen already, coincides with \cite{Gralla:2023oya} (where all the prefactors were provided),
and shows that the mere presence of a horizon in a system induces decoherence on quantum superpositions
with a decoherence functional which is
linearly increasing with the time the superposition has been kept open.
Notice that if we tag decoherence using the function $\cal D$ introduced in (\ref{deco})
(and used in \cite{Danielson:2022tdw,Danielson:2022sga,Danielson:2024yru}), 
we get the relation
$\mathcal{D} = 1-e^{-\frac{1}{2} \mathds{D}}$
which for small $\mathds{D}$ reduces to
${\cal D} \approx \frac{1}{2} \, \mathds{D}$.

As we shall see in section 3, the consideration of quantum properties of the horizon leads to put some limit
on how small the frequencies, which effectively contribute
to the integrals in (\ref{doubleD}), (\ref{doubleD_2}), or (\ref{N_2}), can be.   
In particular, looking at the derivation as presented above, this brings with it to challenge the assumption of the small frequency limit.
The $\bar\Omega \ll a$ should then be released and some effective 
low frequency cut should be inserted instead.
We will proceed then to study 
the expression
(\ref{doubleD}), i.e., the results before the $\bar\Omega \ll a$ limit is applied. 
Since, as emphasized in \cite{Danielson:2024yru},
the very low frequencies in the integration are essentially responsible for the decoherence effect, we might expect that this lower cut in frequency might bring in some significant effects on the decoherence pattern.

Both the integrals (\ref{doubleD_2}) and (\ref{N_2}), and 
(\ref{doubleD}) with greater reason, 
are clearly diverging if the upper integration limit is $\infty$, this being ascribable to the sharpness of the transition to open or close the dipole.
For smooth transitions, they are expected to be converging and this is expressed at an effective level by the introduction of the upper cuts $\bar\omega$ and $\bar\Omega$.
To delve deeper into these aspects, 
and have an explicit example to consider in detail,
in the following subsection this is checked using a specific, analytically convenient form of the transition function.

\subsection{A specific choice for the dipole moment}\label{sec:dipmom}

Let us introduce a specific form of the dipole moment $\epsilon^A(\tau)$ in order to be able to evaluate the decoherence functional \eqref{eq:defdecfctft} a bit more in detail 
and to explicitly see the contribution of the time ${\cal T}_1$ during which the dipole is created and destroyed, 
and also to have control on the influence of the spectrum 
at frequencies other than the $\Omega\to 0$ limit. 

\begin{figure}[!hbtp]
  \centering
  \includegraphics[width=0.33\textwidth]{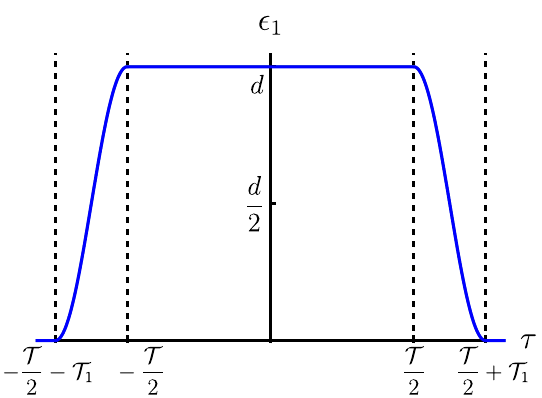}
  \caption{Plot of the dipole moment $\epsilon_1(\tau)$ in equation \eqref{eq:dpmom}.
  }\label{fig:pldm}
\end{figure}

With reference to equation \eqref{eq:dipmomsolx},
we choose
\begin{align}\label{eq:dpmom}
\resizebox{0.95\hsize}{!}{$
    \epsilon_1(\tau) = \begin{cases}
d & \text{for } -\frac{\cal T}{2} \leq \tau \leq \frac{\cal T}{2}\\
\frac{d}{2} \left(1+ \cos\left[ \frac{\tau-\frac{\cal T}{2}}{{\cal T}_1} \pi\right] \right) & \text{for } \frac{\cal T}{2} < \tau < \frac{\cal T}{2}+{\cal{T}}_1\\
\frac{d}{2} \left(1+ \cos\left[ \frac{\tau+\frac{\cal T}{2}}{{\cal T}_1} \pi\right] \right) & \text{for } -\frac{\cal T}{2}-{\cal T}_1 < \tau < -\frac{\cal T}{2}\\
0 & \text{else\,,}
\end{cases}
$}
\end{align}
the other components of $\epsilon_A$ are assumed to vanish.
The form of this dipole moment is visualized in figure \ref{fig:pldm} and is chosen such that $\epsilon_1(\tau) \in C^1(\mathds{R})$.

Fourier transforming to $\tilde\epsilon_1$ we
get (the details are reported in Appendix \ref{sec_App0}):
\begin{align}\label{doubleD_transients}
    \mathds{D} &= \frac{2}{\pi} \ C \int_{0}^\infty d\Omega\;  \frac{a^2+\Omega^2}{\Omega}  \frac{\cos^2\left(\frac{{\cal T}_1}{2}\Omega\right)}{\left(1-\frac{\Omega^2 {\cal{T}}_1^2}{\pi^2}\right)^2} \nonumber\\ &\hspace{1.2in}\cdot\sin^2\left(\frac{{\cal T}+{\cal T}_1}{2}\Omega\right) \coth\left(\frac{\pi\Omega}{a}\right)\,,
\end{align}
which is finite indeed, as expected.
The Appendix also shows that this integral, in the large $\cal T$ limit,
reduces to the final expression in equation \eqref{eq:Dfctapro1}.

This concludes the discussion of the decoherence induced by a classical horizon, which was derived in \cite{Danielson:2022tdw,Danielson:2022sga,Gralla:2023oya,Wilson-Gerow:2024ljx,Danielson:2024yru} and which we complemented here with a specific choice of dipole moment, that will be of use for some particular aspects
when considering a 
horizon with quantum geometry
in the next section.

\section{Decoherence in the presence of horizons with quantum 
geometries}\label{sec:quantum}

So far, we have discussed the decoherence caused in the particle's superposition due to the presence of a
horizon whose geometry is treated classically
--still the black hole behaving 
as a quantum system \cite{Biggs:2024dgp}--
along the lines of \cite{Danielson:2022sga,Danielson:2022tdw,Gralla:2023oya,Wilson-Gerow:2024ljx,Danielson:2024yru}. 
All these works agree on the fact that the decoherence, which turns out to be linear with the time which the superposition has been kept open, 
is caused by the low frequency modes of the field piercing the horizon.
Our aim is now to see how this description gets modified 
in case quantum aspects of the geometry itself of the horizons are taken into account, specifically in the form
of the existence of quantum geometries
with discrete energy levels.

The element of quantumness of geometry we go to consider
is the existence, at least at an effective
level, of a fundamental limit length,
below which, operationally, quadratic intervals between
events do loose meaning.
This is one of the most basic expectations from attempts 
to combine the main tenets of general relativity with quantum field theory,
and is indeed something foreseen by most, 
wildly diverse, approaches to quantum gravity 
(see \cite{Garay:1994en} and references therein).
It goes clearly with quite a bold assumption,
since it points to length scales so tiny
(typically the Planck length) that are far beyond
what is directly observable right now
and for which we have no assurance that physics
itself still works according to the rules
we know from much larger scales.

The limit length acts as a regulator of divergences
in quantum field theory and of singularities 
in general relativity \cite{Deser:1957zz}.
DeWitt brought this to the level to write
a quantum-gravity effective action, 
which removes the singularity, 
and in which the propagator gets a form corresponding
to add to the quadratic distance a constant
interpretable as a residual distance when
the two points go to coincide 
\cite{dewitt1964gravity, dewitt1981approximate}.
The meaning of this distance as a zero-point length has been further explored
by Padmanabhan in \cite{padmanabhan1985physical}.

Apparently there
are not so many options available
in the description of the smallest scales 
when starting from general relativity and quantum mechanics at reasonable scales.
Efforts, for example, to violate or to modify Lorentz invariance 
at small scales, are severely constrained by experimental data
or go into conceptual issues,
giving it more likely to give up with locality rather than Lorentz invariance
when at the very small scales 
\cite{garay1998spacetime, Doplicher:1994tu, Sorkin:2007qi, giddings2006locality}.

In view of the above,
a metric description \cite{KotE, KotF, KotI}
has been explicitly developed in recent years,
as mentioned in the Introduction,
which effectively incorporates in a Lorentz-invariant manner
the existence of a limit length.
It does so
through a modification of the quadratic distance,
an inherently nonlocal quantity
(it is a biscalar, it depends on two points),
to a new quadratic distance
whose nonvanishing in the coincidence limit
between the points
exhibits nonlocality in the smallest scales. 

A key prediction of this
new metric is that
volumes in the space transverse
to the geodesic connecting the
points go to a finite limit at coincidence instead of vanishing \cite{Padmanabhan:2015pza}.
When this is applied to null geodesics, and specifically to
horizon generators, it gives a finite area limit in the transverse space \cite{QuantumMetricNull}, and then a finite area gap for changes of horizon area,
with, as incredible as might seem
as we are dealing possibly with Planck scale effects, ensuing experimental features even potentially detectable in the future with gravitational wave antennas
\cite{Cardoso:2019, Agullo:2021, KriPer}.

We focus 
then
on a situation in which the black hole
is in a certain state with area
$A$ and consider the possible change of
area to the next allowed larger value
$A' = A + A_0$
due to some absorptive process from
its surroundings
with $A_0$ the limit area mentioned above.
By conservation of energy, one expects that a transition like this is possible only if the black hole absorbs enough energy $E$ to fill the energy gap
between the masses corresponding to the two different states of the black hole: 
$E \ge M' - M \equiv E_0$.

For a Schwarzschild black hole of (initial) mass $M$,
which is the case we focus on in this paper,
$E_0$ is related to the quantum of area $A_0$
by $E_0 = \frac{\kappa}{8\pi} A_0$, with 
$\kappa = \frac{1}{4 M}$ the surface gravity.
For the quantum of area,
the minimum-length metric 
foresees 
\cite{QuantumMetricNull, KriPer}
$A_0 = 4\pi L^2 = 4\pi \beta^2 l_p^2$, with $\beta \equiv \frac{L}{l_p}$ the ratio of 
minimal length $L$ to the Planck length $l_p$.
When putting here $\beta = \sqrt{2}$ we recover 
for $A_0$
the quantum of area found in \cite{Bekenstein} and \cite{Maggiore};
when we allow for $\beta \ll 1$, we move towards the results contemplated in \cite{Barreira, Barbero}
(where the area spacing decreases when mass increases)
with an almost continuous variability of the area for solar mass black holes, akin to the classical-geometry case. 

Conservation of energy requires that there is a threshold frequency
\begin{eqnarray}
\omega_0 = E_0 = \frac{\kappa}{2} \beta^2 l_p^2,
\end{eqnarray}
in asymptotic time, for modes to be absorbed by the black hole.

When the same argument is applied to the vacuum,
using as vacuum state the Unruh vacuum 
\cite{Unruh}
on the past horizon, this goes with that the thermally populated vacuum we have with respect to Killing time gets depleted of modes with $\omega < \omega_0$;
indeed, these modes do not have enough energy to compensate for the minimum shrinking of the area.
Our assumption, actually
the key assumption of the paper,
is then the following:
In the expansion of a field 
impinging on the horizon,
only modes with frequency $\omega > \omega_0$
can be absorbed by the horizon
and then can determine the decoherence functional.
In practice this means that in the computation of the decoherence as given in equation (\ref{N_2}) one ought to insert the cut $\omega > \omega_0$. For equations (\ref{doubleD}) or (\ref{doubleD_transients})
this amounts to 
$\Omega > \Omega_0 = \frac{a}{\kappa} \, \omega_0.$

The physical picture we are maintaining here is that, due to the area gap, not all the modes of the field produced in the particle's lab can reach beyond the horizon,
in particular those with asymptotic frequency small can not. 
Reading this in terms of information that can go
beyond the horizon
it means that maybe it can turn out quite small 
and an observer there can not, using it, discriminate 
the path of the particle's superposition.
This would imply no need to have the particle decohering 
to have peaceful coexistence of complementarity and causality.

We will
proceed working with a Rindler horizon, regarded as the approximating Rindler horizon of the actual bifurcate Killing horizon of the black hole in the near-horizon limit.
But,
due to the equivalence noted right below Eq. (\ref{doubleD_2}),
the results we will obtain 
from existence of a frequency cut
clearly extend essentially
unaltered to
(\ref{N_2}) in the global description
(with surface gravity $\kappa$ replacing the local acceleration $a$ in the integral,
as well as the asymptotic time and frequency, $T$ and $\omega$, replacing the local $\mathcal{T}$ and $\Omega$),
for which no restriction is present on the distance to the horizon.   
In section \ref{sec:qualdisc} we present a qualitative analysis on the impact of this effect on the decoherence functional. In section \ref{sec:quhres1} we then perform the corresponding analytical calculation.

\subsection{Qualitative discussion}\label{sec:qualdisc}

By imposing a cut in the frequency integration that excludes frequencies below the one absorbable by the horizon $\Omega_0$, the decoherence functional (\ref{doubleD}) becomes
\begin{equation}\label{doubleD_cut}
    \mathds{D} = \frac{2}{\pi a^2}  \ C \int_{\Omega_0}^{\bar\Omega} d\Omega  \ \frac{a^2 + \Omega^2}{\Omega} \sin^2\left(\frac{\Omega {\cal T}}{2}\right) \coth\left(\frac{\pi \Omega}{a}\right)\
\end{equation}
or explicitly, in terms of the transient time of our specific model in equation \eqref{doubleD_transients},
\begin{align}\label{eq:decfctC}
    \mathds{D} &= 
    \frac{2}{\pi a^2} \ C
    \int_{\Omega_0}^\infty d\Omega\;  \frac{a^2+\Omega^2}{\Omega}  \frac{\cos^2\left(\frac{{\cal T}_1}{2}\Omega\right) }{\left(1-\frac{\Omega^2 {\cal T}_1^2}{\pi^2}\right)^2} \nonumber\\ &\hspace{1.2in}\cdot \sin^2\left(\frac{{\cal T}+{\cal T}_1}{2}\Omega\right)  \coth\left(\frac{\pi\Omega}{a}\right)\nonumber \\ & =: \frac{2}{\pi a^2} C \int_{\Omega_0}^\infty d\Omega\; I_{\mathcal{T}}(\Omega)\,,
\end{align}
where, following the original idea of the gedankenexperiment, we focus on the case
${\cal T}_1 \ll {\cal T}$. For later reference, we defined the integrand function $I_\mathcal{T}(\Omega)$.

Most of the analysis which will follow is essentially the same for both 
the expressions \eqref{doubleD_cut} and
\eqref{eq:decfctC}.
We would like however to put some emphasis on those features which
do hinge on the transients,
and see, in our (fairly generic and mathematically tractable) example, how they arise.
We refer then, in the discussion
below, to the explicit expression \eqref{eq:decfctC}.

\subsubsection{Analysis of the integrand}\label{sec:anofint}
Let us start with a first 
approximate analysis to get an idea of the qualitative changes the cut implies.
Let us consider the integrand $I_{\cal T}(\Omega)$ in \eqref{eq:decfctC}.
In the limit of vanishing $\Omega$ we find
\begin{equation}\label{eq:ITWat0}
    \lim_{\Omega\to 0} I_{\cal T}(\Omega) = \frac{a^3}{12 \pi^3} ({\cal T}+{\cal T}_1)^2 \approx \frac{a^3}{12 \pi^3} {\cal T}^2\,,
\end{equation}
hence the height of the integrand at $\Omega=0$ grows as ${\cal T}^2$.
Assuming ${\cal T}\gg {\cal T}_1$, the first zero of the integrand is at
\begin{equation}
    \Omega\approx \frac{2\pi}{\cal T}
\end{equation}
and we already see here that the integral,
if taken from $\Omega = 0$ to the first zero, would scale as $\sim {\cal T}$, and the presence of the frequency cut can clearly be expected to affect this result.
For larger values of $\Omega$, the amplitude is completely determined by the enveloping function
\begin{equation}\label{eq:envelfct}
    E(\Omega):=\frac{a^2+\Omega^2}{\Omega}  \frac{1}{\left(1-\frac{\Omega^2 {\cal T}_1^2}{\pi^2}\right)^2}  \coth\left(\frac{\pi\Omega}{a}\right)\,,
\end{equation}
which is independent of $\cal T$. 

\begin{figure*}[!tbp]
  \centering
  \includegraphics[width=0.45\textwidth]{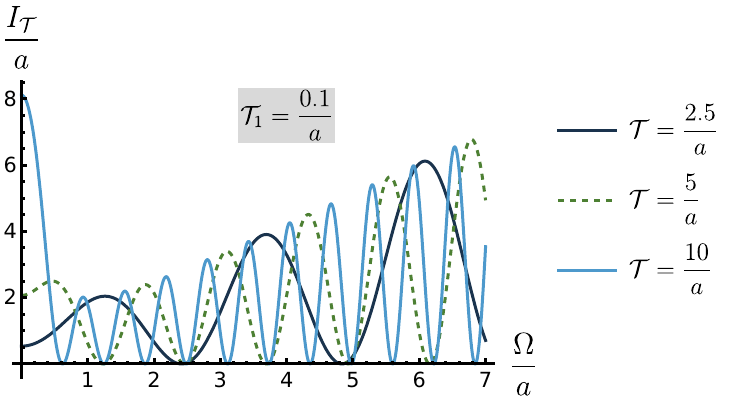}
  \hfill
  \includegraphics[width=0.45\textwidth]{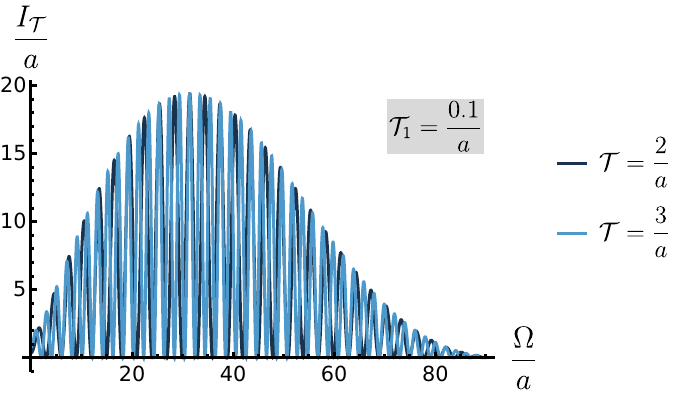}\\
  \includegraphics[width=0.45\textwidth]{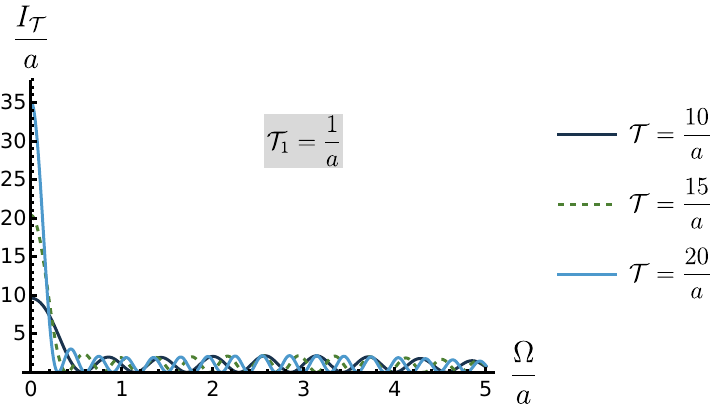}
  \hfill
  \includegraphics[width=0.45\textwidth]{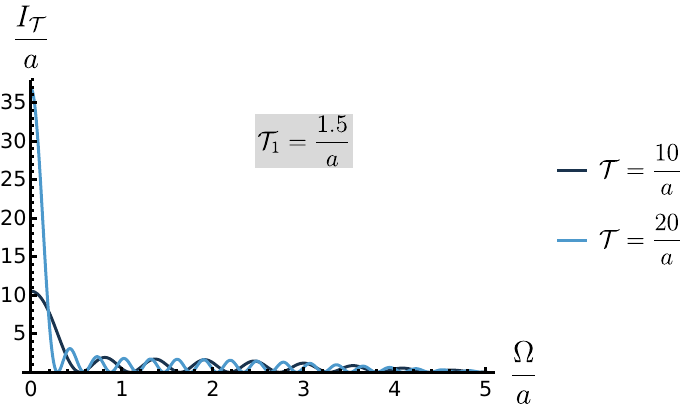}
  \caption{Visualization of the integrand $I_T(\Omega)$ defined in equation \eqref{eq:decfctC} in units of $a$ for different values of $a \mathcal{T}_1$ and $a \mathcal{T}$ such that $\mathcal{T}\gg \mathcal{T}_1$. It can be seen in particular from the top left plot that the oscillations are determined by the enveloping function $E(\Omega)$ defined in equation \eqref{eq:envelfct} and are approximately independent of $\mathcal{T}$ after the first zero. The different values of $\mathcal{T}$ manifest in particular in the value of $I_\mathcal{T}(\Omega)$ at $\Omega=0$, which is proportional to $\mathcal{T}^2$. As the enveloping function changes slowly for $ \mathcal{T}\gg \frac{1}{a}$ compared to the oscillations, the difference of the integral, that is of the area of $I_\mathcal{T}(\Omega)$ for different values of $\mathcal{T}$ is mostly determined by the behavior before the first zero. With increasing $\mathcal{T}$, the oscillations become faster (i.e. the period in $\frac{\Omega}{a}$ reduces). In the top right plot one can see that $\mathcal{T}_1>0$ acts as a UV-regulator that damps the integrand for $\Omega \mathcal{T}_1 \gg 1$. The plots in the second line show the influence of changing $a\mathcal{T}_1$, where larger values in $\mathcal{T}_1$ lead to a quicker damping of the enveloping function. }\label{fig:plInt}
\end{figure*}

Let us now consider the behavior of the integrand $I_{\cal T}(\Omega)$ integrated over the next oscillation period $\Omega \in \left[\frac{2\pi}{\cal T},\frac{4\pi}{\cal T}\right]$ for some large $\cal T$ that fulfills $a {\cal T} \gg 1$. 
Given additionally the condition ${\cal T}\gg {\cal T}_1$, the envelope function $E(\Omega)$ will not change fast on this interval. 
Hence for a larger $\widetilde{{\cal T}} > {\cal T}$, the integral of $I_{\widetilde{\cal T}}(\Omega)$ on the same interval $\Omega \in \left[\frac{2\pi}{\cal T},\frac{4\pi}{\cal T}\right]$ will yield approximately the same result, independently of $\cal T$ or $\widetilde{\cal T}$. 

Thus for $\Omega > \frac{2\pi}{\cal T}$ the integration gives approximately a constant contribution to the decoherence functional, independent of $\cal T$. Here, the inclusion of a non-zero time ${\cal T}_1$ is crucial to obtain a finite value for this offset in the decoherence functional, as in the limit ${\cal T}_1\to 0$ the integrand $I_{\cal T}(\Omega)$ goes linear in $\Omega$ for large $\Omega$ and hence would lead to a UV-divergent decoherence function.

The first period of the integration yields
\begin{equation}
    I_{\cal T}(\Omega) \approx \frac{a^3}{\pi \Omega^2} \sin^2\left(\frac{\Omega {\cal T}}{2}\right)
\end{equation}
and thus the decoherence caused by this part without a cut can be approximated as
\begin{equation}
    \frac{2 a}{\pi^2} \, C
    \int_0^{\frac{2\pi}{\cal T}} d\Omega\; \frac{1}{\Omega^2} \sin^2\left(\frac{\Omega {\cal T}}{2}\right) 
    = \frac{a}{2\pi}\, C \  \frac{\text{SI}(2\pi)}{\pi/2} \, {\cal T}\,,
\end{equation}
which gives a linear dependence on $\mathcal{T}$,
with coefficient close to that given in \eqref{eq:Dfctapro1}
and found in \cite{Gralla:2023oya, Wilson-Gerow:2024ljx}.
Here, $\text{SI}(x)$ denotes the sine integral function, which is $\text{SI}(2\pi) \approx 1.42$ and approaches $\frac{\pi}{2}$ for arguments much larger than one.

Figure \ref{fig:plInt} displays the integrand $I_{\cal T}(\Omega)$ in some
ranges of $\Omega$
for various choices of $\cal T$ and ${\cal T}_1$ expressed as functions of $a$.

\subsubsection{Estimate of decoherence}\label{sec:festdecqh}

These discussions show that,
for each choice of $\cal T$,
we can recognize two different contributions to decoherence:
one, from the region $\Omega < \frac{2\pi}{\cal T}$, giving
a linear increase with $\cal T$,
and the complementary one, $\Omega > \frac{2\pi}{\cal T}$,
giving a constant offset (of value determined by ${\cal T}_1$).
If we now implement a low frequency cut $\Omega_0$, 
then for any 
time ${\cal T} > \frac{2\pi}{\Omega_0}$, 
the first period of the oscillation is excluded 
from our integration domain, 
hence the linear dependence of the decoherence functional on $\cal T$ has to turn into a constant value, independent of $\cal T$, which we refer to as saturation value.

To foster this discussion and see the qualitative difference brought in even by a
very small cut $\Omega_0 \ll a$, $\Omega_0 {\cal T}_1 \ll 1$, let us choose a second, auxiliary frequency $\Omega'$ such that
\begin{equation}\label{constraints}
    \Omega' > \Omega_0 \hspace{0.5in} \Omega' {\cal T}_1 \ll 1 \hspace{.5in} \Omega' \ll a\,.
\end{equation}

Our interest is to study the situation for increasing $\cal T$.
We choose $\cal T$ such that 
$\Omega' {\cal T} \ge 2\pi$,
which means that integration till $\Omega'$ includes the first oscillation period. 
From the preceding discussion we know that this is the region where  the dependence of $\mathds{D}$ on 
${\cal T}$ should
not be constant in general,
and that the integration on the domain $\Omega > \Omega'$
gives a
constant quantity, independent of $\cal T$.
Writing $\mathds{D} = \mathds{D}' + {\rm const}$ 
we have then,
assuming as above ${\cal T}\gg {\cal T}_1$:
\begin{align}\label{eq:ap1wc}
    \mathds{D}' &\approx
\frac{2 a}{\pi^2} \, C \,
\int_{\Omega_0}^{\Omega'} d\Omega \frac{\sin^2\left( \frac{\cal T}{2}\Omega \right)}{\Omega^2} \nonumber\\&= 
\frac{2 a}{\pi^2} \, C \,
\Bigg[ \frac{\Omega' \,{\cal T} \,\text{SI}(\Omega'{\cal T}) + \cos(\Omega'{\cal T}) - 1  }{2\Omega'}  \nonumber\\ &\hspace{0.9in}- \frac{{\Omega}_0\, {\cal T}\, \text{SI}(\Omega_0 {\cal T}) + \cos(\Omega_0 {\cal T}) - 1  }{2 \Omega_0}\Bigg]\,.
\end{align}
In this regime, $\Omega' \, {\cal T} \, \text{SI}(\Omega' {\cal T}) \approx \frac{\pi}{2} \Omega'\, {\cal T} - \cos(\Omega' {\cal T})$, thus we can approximate
\begin{equation}
\resizebox{1.\hsize}{!}{$
\mathds{D}' \approx  
\frac{2 a}{\pi^2} \, C \,
\left[ \frac{\pi}{4} {\cal T} - \frac{1}{2 \Omega'}- \frac{{\Omega}_0\, T\, \text{SI}(\Omega_0 {\cal T}) + \cos(\Omega_0 {\cal T}) - 1  }{2 \Omega_0}\right].
$}
\end{equation}

For $\Omega_0$ small enough to have
$\Omega_0 {\cal T} \ll 1$, 
we can use 
\begin{align}
{\Omega}_0\, {\cal T}\, \text{SI}(\Omega_0 {\cal T}) + \cos(\Omega_0 {\cal T}) - 1  &\approx \Omega_0^2 {\cal T}^2 +1 - \frac{1}{2} \Omega_0^2 {\cal T}^2 +1 \nonumber\\ &= \frac{1}{2} \Omega_0^2 {\cal T}^2,    
\end{align} 
which yields
\begin{equation}\label{smallOmega0Tlimit}
    \mathds{D}' \approx  
    \frac{2 a}{\pi^2} \, C \, \, 
    \frac{1}{4} \Big(\pi - \Omega_0 {\cal T}- \frac{2}{{\Omega'}{\cal T}}\Big) \, {\cal T} \approx \frac{a}{2 \pi} \, C\,  {\cal T}  \,,
\end{equation}
where in the last step we assumed $\cal T$ large enough to have $\Omega' {\cal T} \gg 1$.
Hence we get, as expected, the same behavior as in the case without cut discussed above in section \ref{sec:ClasHor}. 

However, once, for fixed $\Omega_0$, ${\cal T}$ increases such that $\Omega_0 {\cal T} \geq 2\pi$, we can approximate ${\Omega}_0\, {\cal T}\, \text{SI}(\Omega_0 {\cal T}) + \cos(\Omega_0 {\cal T}) - 1  \approx \frac{\pi}{2} \Omega_0 {\cal T} -1$ and obtain
\begin{equation}\label{eq:qualanalfirstDec}
    \mathds{D}' \approx  \frac{a}{2 \pi} \, C \, 
    \left[ \frac{2}{\pi\Omega_0} - \frac{2}{\pi{\Omega'}}\right] \approx 
    \frac{a}{\pi^2} \, C \, 
    \frac{1}{\Omega_0}
    \equiv \mathds{D}_{\text{sat}} \,,
\end{equation}
where in the last step
we still assumed ${\Omega'} \gg \Omega_0$. 
The dependence on $\cal T$ disappears (as the first oscillation cycle lies now to the left of
the lower limit of integration) and the contribution comes from the terms outside the first oscillation cycle
which are approximately the same for any $\mathcal{T}$. 
This saturating value, $\mathds{D}_{\text{sat}}$, adds in principle to the constant coming from integration beyond $\Omega'$;
as it will be discussed later however, by a suitable choice of the transients (i.e. creation and destruction of the dipole)
the latter contribution can be shown to be negligible, i.e., $\mathds{D}' = \mathds{D} = \mathds{D}_{\text{sat}}$, when $\Omega_0 \ll a$.
The same, i.e., $\mathds{D}' = \mathds{D}$ in the same limit, can be said also for Eq. (\ref{smallOmega0Tlimit}) for $\cal T$ not too small.

The expansion for large argument of $\text{SI}(\Omega_0 \mathcal{T})$ yields, next to the constant factor $\frac{\pi}{2}$, oscillatory terms which come with inverse powers of the argument, hence are damped for increasing $\mathcal{T}$. The above suggests that the saturation value depends inversely on the cut in the regime considered here.

Once we consider larger cuts,
like $\Omega_0 \approx a$ or larger,
we can not, of course, introduce an auxiliary frequency $\Omega'$
along the lines above.
Also, we cannot use any more for the integrand the limit expression for $\Omega\to 0$,
but we have to resort to the full expression in 
\eqref{eq:decfctC}.
For the same times $\cal T$ considered above, however,
we clearly have $\Omega_0 {\cal T} \ge 2\pi$.
Considering the integration on all the domain $\Omega \ge \Omega_0$
we have then the first oscillation completely to the left of the lower limit of integration. We already know then that the integral can not depend on $\cal T$, i.e., for these values of $\cal T$ we are already
in the constant part, whose value will depend on $\Omega_0$ as above, and on the chosen ${\cal T}_1$.
Overall, if $\Omega_0 \geq a$, the linear dependence on $\mathcal{T}$ is lost at times (much) smaller than when $\Omega_0 \ll a$.

\subsubsection{Estimates for saturation values and transition times}\label{sec:svalandttime}

From the previous discussion we see that there exists
a critical time ${\cal T}_{\rm crit} \sim \frac{1}{\Omega_0}$ at which $\mathds{D}$ switches from a linear increase with $\cal T$ to a constant behavior 
independent of $\cal T$, sort of saturating value $\mathds{D}_{\rm sat}$, given by Eq. (\ref{eq:qualanalfirstDec}).
Let us take then ${\cal T}_{\rm crit}$ as the time
the separation has been kept open to provide 
$\mathds{D} = \mathds{D}_{\text{sat}}$ under linear increase from $\mathds{D} = 0$ at ${\cal T} = 0$,
i.e., 
${\cal T}_\text{crit} = \mathds{D}_{\text{sat}} / (\frac{a}{2\pi} C) = \frac{2}{\pi} \frac{1}{\Omega_0}$,
where we use for the linear increase in $\cal T$
the coefficient reported in \cite{Gralla:2023oya, Wilson-Gerow:2024ljx}, which we also obtain in the small $\Omega_0$, not too large $\cal T$, limit (cf. Eq. (\ref{smallOmega0Tlimit})).

This can be compared with the decoherence time were the linear increase in $\mathds{D}$ kept forever.
Writing the Feynman-Vernon functional in (\ref{deco_2}) as
$|{\cal F}| = e^{-\frac{\cal T}{{\cal T}_{\rm dec}}}$,
with ${\cal T}_{\rm dec} = \frac{4 \pi}{a C}$ the characteristic decoherence time, at which $\mathds{D} =2$,
we clearly get 
$\frac{{\cal T}_\text{crit}}{{\cal T}_\text{dec}}
= \mathds{D}_{\text{sat}}/2$ so that, of course,
any value of $\mathds{D}_{\text{sat}}$ gives also the order
of magnitude of the ratio  of the critical to the decoherence time.

We now use the fact that near the horizon 
the acceleration $a$ of the particle's lab is related to the proper distance $D$ to the horizon by 
$a= \frac{1}{D} + {\cal O}(D)$,
to cast $\mathds{D}_\text{sat}$ in equation (\ref{eq:qualanalfirstDec})
in the form
\begin{equation}\label{eq:Dsataprox}
    \mathds{D}_{\rm sat} \approx
    \frac{q^2}{3 \pi^3 \epsilon_0} \Big(\frac{d}{D}\Big)^2 
    \frac{1}{\Omega_0/a} \approx
    \frac{q^2}{3 \pi^3 \epsilon_0} \Big(\frac{d}{D}\Big)^2
    \frac{1}{\omega_0/\kappa}\,,
\end{equation}
where we have used the expression of $C$ implied
by the last line of (\ref{eq:Dfctapro1}).
Taking $q = e$, the electron charge,
this gives
\begin{equation}
  \mathds{D}_{\rm sat}  \approx
  0.98 \cdot 10^{-3} \, \, \, \Big(\frac{d}{D}\Big)^2
    \frac{1}{A_0/(8\pi l_p^2)},
\end{equation}
with 
all the constants reinserted.

\begin{figure}[!htbp]
  \centering
  \includegraphics[width=0.45\textwidth]{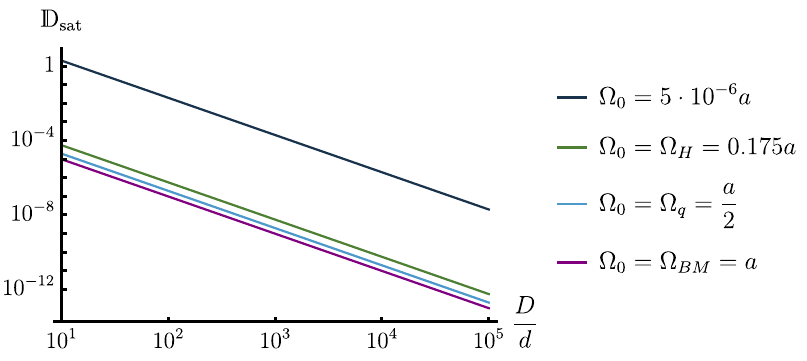}
  \caption{
  Dependence of the saturation value in equation \eqref{eq:Dsataprox} on $\frac{D}{d}$ for different cutoff values $\Omega_0$. 
  The lowest value $\Omega_0 = 5\cdot 10^{-6} a$ was chosen such that $\mathds{D}_{\rm sat} \approx 2$ for $\frac{D}{d}=10$.   }\label{fig:plDsat} 
\end{figure}

This relation shows that 
what matters in setting the saturation value of the decoherence is the ratio $D/d$
between distance and separation of the superposed positions, and the size of the minimum area change $A_0$
to have absorption, with $\mathds{D}_{\rm sat}$ decreasing
when the two increase.
Fig. \ref{fig:plDsat} displays $\mathds{D}_{\rm sat}$
as a function of $D/d$ for selected values of $A_0$,
proposed in the literature.
It is evident from the figure that for them the effect
of the frequency cut 
is to reduce the maximum decoherence 
one can get due to the presence of the horizon
to negligible values, even when going very near to the 
horizon.
This means that, in case there is a minimum value for the area changes of the order of magnitude reported in literature, 
the presence of the horizon 
would not be able to induce appreciable decoherence, no matter how long we take open the superposition. 
The figure also shows, as reference, the results for a value of $A_0$ small enough to allow
for a reasonably large value of $\mathds{D}_{\rm sat}$
at least when approaching the horizon 
(taken this here to mean $D \approx 10 \, d$,
just to have a lower limit $D$ compatible
with the assumption $d\ll D$).

\subsubsection{Visualization}\label{sec:plotsdeco}

To conclude this section we explicitly calculate 
to some accuracy level
the full expression of decoherence functional $\mathds{D}$
of Eq. (\ref{eq:decfctC}) as a function of $\cal T$
for different choices of the frequency cut $\Omega_0$.
We work first in the assumption, shared by
\cite{Gralla:2023oya, Wilson-Gerow:2024ljx},
that the transients can be chosen such that
the integral we have to consider
is, at an effective level, 
up to a finite value $\bar\Omega$ as in (\ref{doubleD}).
As it will be clear 
from inspection of the results we will get,
an adequate choice is to take $\bar\Omega = a$.
In our expression with explicit transients
(\ref{eq:decfctC}) this amounts to integrate
up to $\bar\Omega = a$ and to neglect ${\cal T}_1$,
that is to assume ${\cal T}_1 \ll {\cal T}$ (which we always assume in the paper)
and ${\cal T}_1 \, a \ll 1$.
The effect of violating the last condition, namely to consider
${\cal T}_1 \, a \gtrsim 1$, in the integral up to $\bar\Omega$
is simply to introduce a further modulation, which becomes a factor $\frac{1}{2}$ when 
${\cal T}_1 \, a \gg 1$.
In the final part of this subsection we will explicitly
comment on the effects of ${\cal T}_1$ on the integration 
of $I_{\cal T}$ beyond $\bar\Omega$ in (\ref{eq:decfctC}),
and show that for $\Omega_0 \ll a$
and for a suitable choice of ${\cal T}_1$ this last
integration can be consistently ignored.
For $\Omega_0 \approx a$ instead it gives 
a contribution no longer negligible compared to $\mathds{D}_\text{sat}$ obtained before,
but the total decoherence including this part 
is anyway depressed if compared with  $\mathds{D}_\text{sat}$ from smaller $\Omega_0$'s.

In order to obtain analytically solvable integrals 
we approximate the decoherence functional in equation \eqref{eq:decfctC} in the following way:

\begin{align}\label{eq:aprdecfctplt}
    \mathds{D}(\mathcal{T}) \approx &\frac{2}{\pi} \, C \int_{\Omega_0}^{\frac{a}{\pi}} d\Omega\; \frac{\Omega^2 + a^2}{\Omega a^2} \left(\frac{a}{\pi\Omega}+\frac{\pi\Omega}{3a}\right) \sin^2\left( \frac{\Omega \mathcal{T}}{2}\right) \nonumber\\ &+ \frac{2}{\pi}  \, C\int_{\frac{a}{\pi}}^{a} d\Omega\; \frac{\Omega^2 + a^2}{\Omega a^2} \left( 1+2e^{-2\frac{\pi\Omega}{a}}\right) \sin^2\left(\frac{\Omega \mathcal{T}}{2} \right),
\end{align}
where we neglected $\mathcal{T}_1$, and approximated $\coth(x)$ in two different ways: for small argument as $\frac{1}{x}+ \frac{x}{3}$, and for arguments equal to or larger than one as $1+2 e^{-2x}$. This can be estimated to leave the integration result accurate at $<$ 4\% level. 

We then consider times larger than $\approx 1/a$ which guarantees that the integration interval goes beyond the first oscillation of the integrand, and hence the remaining parts only contribute a constant independent of $\mathcal{T}$ (but depending on the specific choice of the dipole and thereby on $\mathcal{T}_1$).

\begin{figure*}[!htbp]
  \centering
  \includegraphics[width=0.46\textwidth]{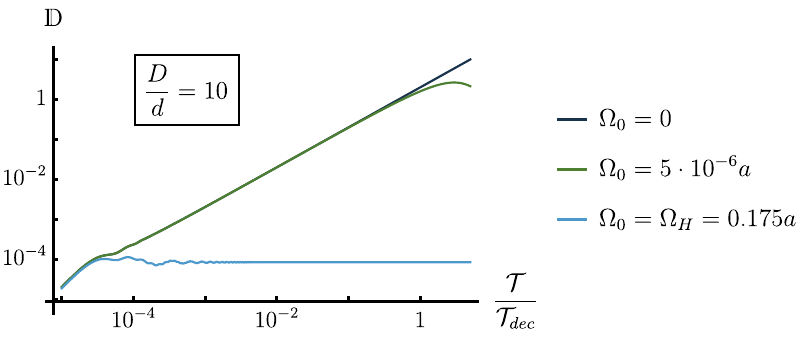}\hfill
  \includegraphics[width=0.46\textwidth]{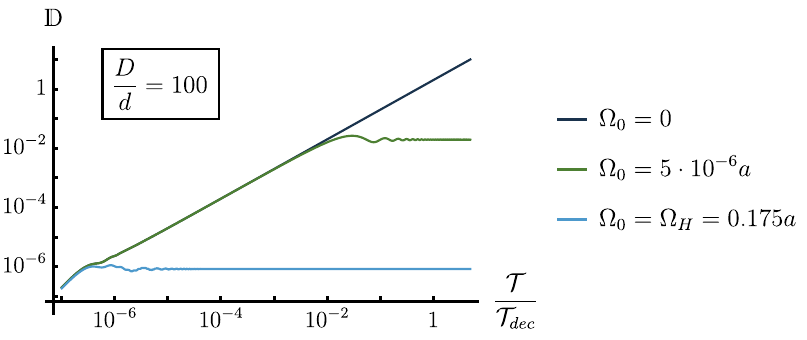}
  \caption{Decoherence functional as a function of $\cal T$ in the approximation of equation \eqref{eq:aprdecfctplt} for different cutoff frequencies using $\frac{D}{d}=10$ (left figure) and $\frac{D}{d}=100$ (right figure). While the case of a classical horizon $\Omega_0=0$ depends linearly on $T$ for large times, for a quantum horizon a saturation is obtained after a certain time.
  For very small $\frac{\mathcal{T}}{\mathcal{T}_{\rm dec}}$ a quadratic increase can be seen (which can be shown to correspond to the first oscillation period extending well beyond the upper integration limit); 
  there then follows a transition period into the linear increase which gets saturated after a certain time depending inversely on the size of the cutoff.
  }\label{fig:plDec} 
\end{figure*}

The effect of a relatively small cut is illustrated in figure \ref{fig:plDec} (choosing $q = e$, as above). 
While for a classical horizon the decoherence grows linearly with $\cal T$ (for large values of $\cal T$)
and therefore crosses 
$\mathds{D}=2$ in ${\cal T}_\text{dec}$
in accordance to \cite{Danielson:2022sga, Gralla:2023oya, Wilson-Gerow:2024ljx}, 
for a quantum horizon, with small but finite value of the cut $\Omega_0$, the linear regime is left after a certain time and the decoherence functional approaches a saturation value with damped oscillations.

As discussed at the beginning of section \ref{sec:quantum}, there are different prospects for values of the asymptotic cutoff frequency $\omega_0$ relating it to the horizon surface gravity $\kappa$ or, which is the same in the near-horizon limit we are considering, of the cutoff frequency $\Omega_0$ to the acceleration $a$ of the locally approximating Rindler frame. 
The corresponding decoherence functions in the approximation in equation \eqref{eq:aprdecfctplt} can be seen in figure \ref{fig:plDec}. 
While for $\cal T$ small there are some oscillations which are dominated by the contribution of the second term in \eqref{eq:aprdecfctplt}, after a certain value of the time $\cal T$ which the separation was kept open,  
the line for a classical horizon ($\Omega_0=0$) starts the linear behavior predicted by the first term in equation \eqref{eq:aprdecfctplt}, while the other lines remain in damped oscillations around a very small saturation value. 

The reason for this is that once the linear increase sets in, which is the result of the first oscillation of the integrand (in the language of section \ref{sec:anofint}), for cuts comparable to $a$ the first oscillation has almost been left and hence only the remaining part of the integrand, which is approximately constant in $\mathcal{T}$, contributes. 
The figure depicts the case of a very small $\Omega_0/a$
and that of $\Omega_H/a$, 
coming from the area step proposed in \cite{Hod}
(and found also in \cite{Barbero2}).
All larger proposed values of the limit area $A_0$, bringing to larger values of $\Omega_0/a$, get further depressed (at the same time requiring some enlargement of the upper integration limit
to $\bar\Omega >a$), and exhibit a saturation value $\mathds{D}_\text{sat}$ which becomes so small to become comparable
with the contribution from the integration above $\bar\Omega$,
which we will consider in a moment.
This justifies the choice $\bar\Omega \sim a$ as upper limit of integration: Point is that the dramatic changes (with respect to the $\Omega_0 = 0$ case) we have seen, happen to start already 
for $\Omega_0$ values way smaller than $a$. 

\begin{figure}[!htbp]\label{fig:fig5}
  \centering
  \includegraphics[width=0.45\textwidth]{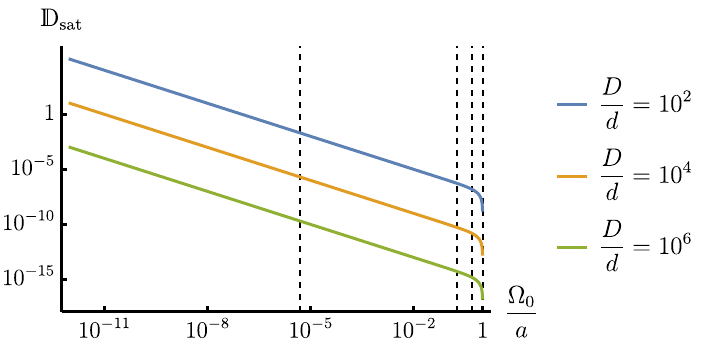}
  \caption{Dependence of the saturation value on the chosen cutoff according to equation \eqref{eq:analresdeco} for different values of $\frac{D}{d}$. For small values of the cutoff, the graph goes as $\Omega_0^{-1}$ while at $\Omega\sim a$ a strong decrease sets in. The vertical lines correspond from left to right to the choices $\Omega_0 = 5\cdot 10^{-6}a$, $\Omega_0 = \Omega_H = 0.175 a$, $\Omega_0 = \Omega_q = \frac{a}{2}$ and $\Omega_0 = \Omega_{BM} = a$.  
  }\label{fig:plDepDO}
\end{figure}

The dependency of the saturation value on the chosen cut $\Omega_0$ can be seen in figure \ref{fig:plDepDO}. As derived in equation \eqref{eq:qualanalfirstDec}, the saturation value depends inversely on the cutoff as long as $\Omega_0 \ll a$. To also access the regime where $\Omega_0 \sim a$ further analysis is needed, and we derive in equation \eqref{eq:analresdeco} below a more general formula which exhibits
a fast decrease of the saturation value once $\Omega_0 \sim a$ is reached. Figure \ref{fig:plDepDO} is based on equation \eqref{eq:analresdeco}.
\\

In the approximation in equation \eqref{eq:aprdecfctplt} the integration interval $\Omega \in [a,\infty]$ is not considered. 
We comment briefly on its contribution
exploiting our specific model for the transients.

In the regime $\Omega > a$, the $\coth$ can be approximated by $1$ and the $\sin$ term averaged due to fast oscillations to give a factor of $\frac{1}{2}$ (with the correction term of order $\frac{\mathcal{T}_1}{\mathcal{T}}$ being neglected as we have $\mathcal{T}_1\ll \mathcal{T}$). 
The additional contribution to the decoherence is then:
\begin{align}\label{eq:t1invis}
    \mathds{D}_{\text{add}} &\approx \frac{C}{\pi a^2} \int_{a}^\infty d\Omega\; \frac{\Omega^2+a^2}{\Omega\left(1- \frac{\Omega^2 \mathcal{T}_1^2}{\pi^2} \right)^2} \cos^2\left( \frac{\Omega \mathcal{T}_1}{2}\right)\nonumber\\
    &= \frac{C}{\pi} \int_1^\infty d\mu \frac{\mu^2 + 1}{\mu \left(1- \frac{16 \mu^2}{C^2} \frac{\mathcal{T}_1^2}{\mathcal{T}_{\rm dec}^2} \right)^2 }  \cos^2\left( \frac{2\pi \mu}{C} \frac{\mathcal{T}_1}{\mathcal{T}_{\rm dec}} \right)\nonumber\\
    &\sim\frac{C}{\pi} \int_1^\infty d\mu \frac{\mu}{ \left(1- \frac{16 \mu^2}{C^2} \frac{\mathcal{T}_1^2}{\mathcal{T}_{\rm dec}^2} \right)^2 }  \cos^2\left( \frac{2\pi \mu}{C} \frac{\mathcal{T}_1}{\mathcal{T}_{\rm dec}} \right)\,,
\end{align}
where in the second line we substituted $\mu := \frac{\Omega}{a}$ and in the third line we used that $\frac{\mu^2+1}{\mu} \sim \mu$ on the integration region we are considering (where $\mu\geq 1$), where $\sim$ means same order of magnitude (one can use as an upper bound for $\mathds{D}_{\rm add}$ twice the expression in the third line).
This can then be integrated analytically and yields, for $\mathcal{T}_1
\ll C \, \mathcal{T}_{\rm dec}$, 
which means ${\cal T}_1 \, a \ll 4 \pi$,
\begin{equation}\label{add1}
    \mathds{D}_{\rm add} \sim \frac{10^{-6}}{16\pi \frac{D^6}{d^6}} \frac{1}{\frac{\mathcal{T}_1^2}{\mathcal{T}_{\rm dec}^2}}\,.    
\end{equation}
On the other hand, 
if ${\cal T}_1$ is large enough to have 
${\cal T}_1 \, a \gg 4 \pi$,
one obtains
\begin{equation}\label{add2}
    \mathds{D}_{\rm add} \sim \frac{10^{-13}}{\pi \frac{D^{10}}{d^{10}}} \frac{1}{\frac{\mathcal{T}_1^4 }{\mathcal{T}_{\rm dec}^4}}\,.
\end{equation}

With reference to Fig. \ref{fig:plDec},
let us choose $\mathcal{T}_1 = 10^{-5} \mathcal{T}_\text{dec}$.
From the positions of the values of $\mathcal{T}_\text{crit}$ we see that
for the $D/d = 10$ case this choice corresponds
to $\mathcal{T}_1 \, a \approx 3.5 \ll 4\pi$.
We use then (\ref{add1}) and we get
$\mathds{D}_\text{add} = \frac{1}{16 \pi} 10^{-2}$.
Note that this value does not depend on the $\Omega_0$ value we consider and is very small in comparison to 1, meaning it has very small effect on total decoherence.
In the figure we see
it results completely negligible compared to
$\mathds{D}_\text{sat}$ from $\Omega_0/a = 5 \cdot 10^{-6}$, but no longer such compared to the saturation
value from $\Omega_0 = \Omega_H$.
It is clear that when $\Omega_0$ 
grows large enough, the corresponding $\mathds{D}_\text{sat}$ becomes small enough
to be comparable or also smaller than the 
$\mathds{D}_\text{add}$ just calculated,
remaining anyway the latter what it is, 
i.e., intrinsically small.

A similar analysis can be performed for the same
choice of $\mathcal{T}_1/\mathcal{T}_\text{dec}$
for $D/d = 100$.
In this case we have 
$\mathcal{T}_1 \,a \approx 350 \gg 4\pi$,
and we have to resort to (\ref{add2}).
We get $\mathds{D}_\text{add} = \frac{1}{\pi} \cdot 10^{-13}$, which is negligible compared to
$\mathds{D}_\text{sat}$ both for $\Omega_0 = 10^{-6} a$
and $\Omega_0 = \Omega_H$ (note however that in the last
case in order to follow the initial linear rise we ought
to take a smaller value of $\mathcal{T}_1)$.
And so on at higher values of $D/d$ with ever decreasing
results for $\mathds{D}_\text{add}$.
Hence, when $D/d$ is small, the contribution to decoherence coming from the time in which the superposition is opened and closed, can be dominant in comparison to the saturation value reached, depending on the size of the cut $\Omega_0$.

The general lesson we learn from this is then as follows.
The influence of this contribution is to add an offset constant in $\mathcal{T}$ to the decoherence functional. As it is additive, it might, depending on the size of $\mathcal{T}_1$, lead to a higher initial value in figure \ref{fig:plDec}. 
The linear increase does then set in only later for higher values of $\mathcal{T}$. This contribution does however not change the qualitative behavior of the decoherence functional for large $\mathcal{T}$ and in particular not the difference between classical and quantum geometry for the horizon.

The qualitative argumentation in this section so far suggests that the effect of the cut is to obtain some saturation of decoherence after a certain time which should be reached before $\mathcal{T} \sim \frac{2\pi}{\Omega_0}$. The saturation value inversely depends on the cutoff value for $\Omega_0\ll a$. In the next section, we solve the decoherence functional analytically for the case of a small cut and discuss the extension to larger values of $\Omega_0$ to give some more insights on its behavior.

\subsection{Analytical calculation}\label{sec:quhres1}
In this section, we analyze in detail the change of behavior of the decoherence functional when a small cut is imposed on the frequencies that can pierce the horizon and thereby check the results from the qualitative discussion in the previous section. To obtain an analytical solution, we make use of techniques from complex analysis. For this, the strategy is to first consider the entire real line as integration domain, that is without a cut, like for a classical horizon, and then study the change effected by the inclusion of a lower frequency cut $\Omega_0$.

\subsubsection{Considering the entire real line as integration domain}\label{sec:analp1}

First,
we consider the decoherence functional 
\eqref{doubleD_transients} expressed, since the integrand is even, as an integral over the entire real line
\begin{align}
    \mathds{D} &= \frac{1}{\pi} \ C \int_{-\infty}^\infty d\Omega\;  \frac{a^2+\Omega^2}{\Omega}  \frac{\cos^2\left(\frac{{\cal T}_1}{2}\Omega\right)}{\left(1-\frac{\Omega^2 {\cal{T}}_1^2}{\pi^2}\right)^2} \nonumber\\ &\hspace{1.2in}\cdot\sin^2\left(\frac{{\cal T}+{\cal T}_1}{2}\Omega\right) \coth\left(\frac{\pi\Omega}{a}\right)\,,
\end{align}
and rewrite the integrand 
by employing the following identity:
\begin{align}\label{eq:defeta}
    &\sin^2\left(\frac{\mathcal{T}+\mathcal{T}_1}{2}\Omega\right) \cos^2\left(\frac{\mathcal{T}_1}{2}\Omega\right)\nonumber\\ &= \frac{1}{4} \text{Re} \left[ 1 - e^{i\Omega(\mathcal{T}+\mathcal{T}_1)} + e^{i\Omega \mathcal{T}_1} - \frac{1}{2} e^{i \Omega \mathcal{T}} -\frac{1}{2} e^{i \Omega (\mathcal{T}+2 \mathcal{T}_1)}\right]\nonumber\\ &=: \frac{1}{4} \text{Re} [\eta(\Omega,\mathcal{T},\mathcal{T}_1)]\,.
\end{align}
As the other terms in the integrand are purely real on the integration interval, we can take the real part after the integration. This leads to
\begin{align}
    \mathds{D} = \frac{C}{4\pi a^2} \text{Re} \int_{-\infty}^\infty d\Omega\; & \frac{a^2+\Omega^2}{\Omega}  \frac{\eta(\Omega,\mathcal{T},\mathcal{T}_1)}{\left(1-\frac{\Omega^2 \mathcal{T}_1^2}{\pi^2}\right)^2} \coth\left(\frac{\pi\Omega}{a}\right).
\end{align}

To perform the integration, the series expansion for the hyperbolic cotangent,
\begin{equation}
    \coth\left( \frac{\pi\Omega}{a}\right) =\frac{a}{\pi \Omega} + \frac{2a}{\pi} \sum_{n=1}^\infty \frac{\Omega}{\Omega^2 + a^2 n^2}\,,
\end{equation}
is applied, which yields
\begin{align}\label{eq:Dtwoterms}
    \mathds{D} =& \underbrace{\frac{C}{4\pi^2 a} \text{Re} \int_{-\infty}^\infty d\Omega\;  \frac{a^2+\Omega^2}{\Omega^2}  \frac{\eta(\Omega,\mathcal{T},\mathcal{T}_1)}{\left(1-\frac{\Omega^2 \mathcal{T}_1^2}{\pi^2}\right)^2}}_{=:\mathds{D}_a}\nonumber\\ &\underbrace{+ \frac{C}{2\pi^2 a} \text{Re} \sum_{n=1}^\infty \int_{-\infty}^\infty d\Omega\;  \frac{a^2+\Omega^2}{a^2 n^2 + \Omega^2}  \frac{\eta(\Omega,\mathcal{T},\mathcal{T}_1)}{\left(1-\frac{\Omega^2 \mathcal{T}_1^2}{\pi^2}\right)^2}}_{=:\mathds{D}_b} \,.
\end{align}

\begin{figure*}[!tbp]
  \centering
  \includegraphics[width=0.45\textwidth]{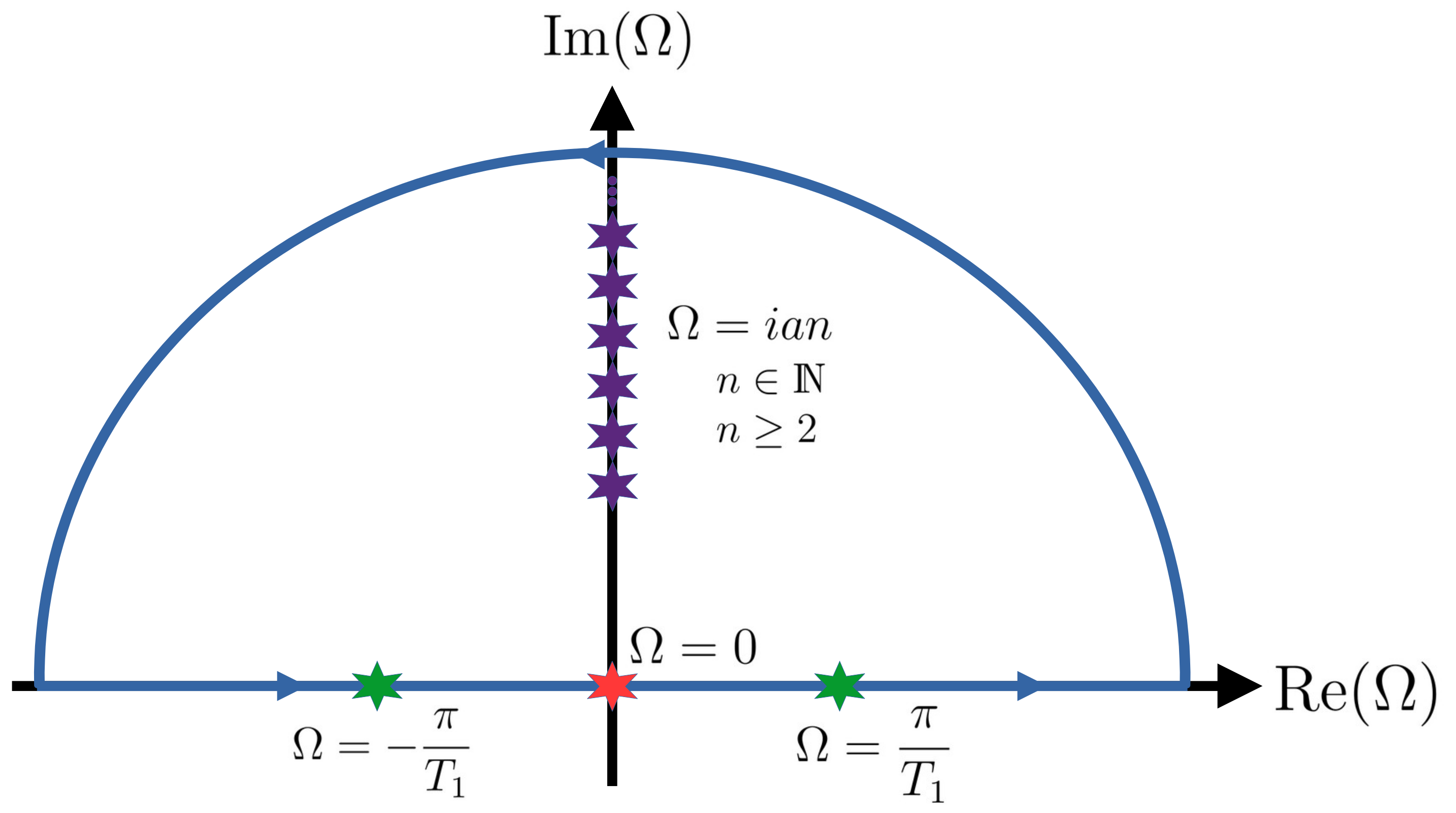}
  \hfill
  \includegraphics[width=0.45\textwidth]{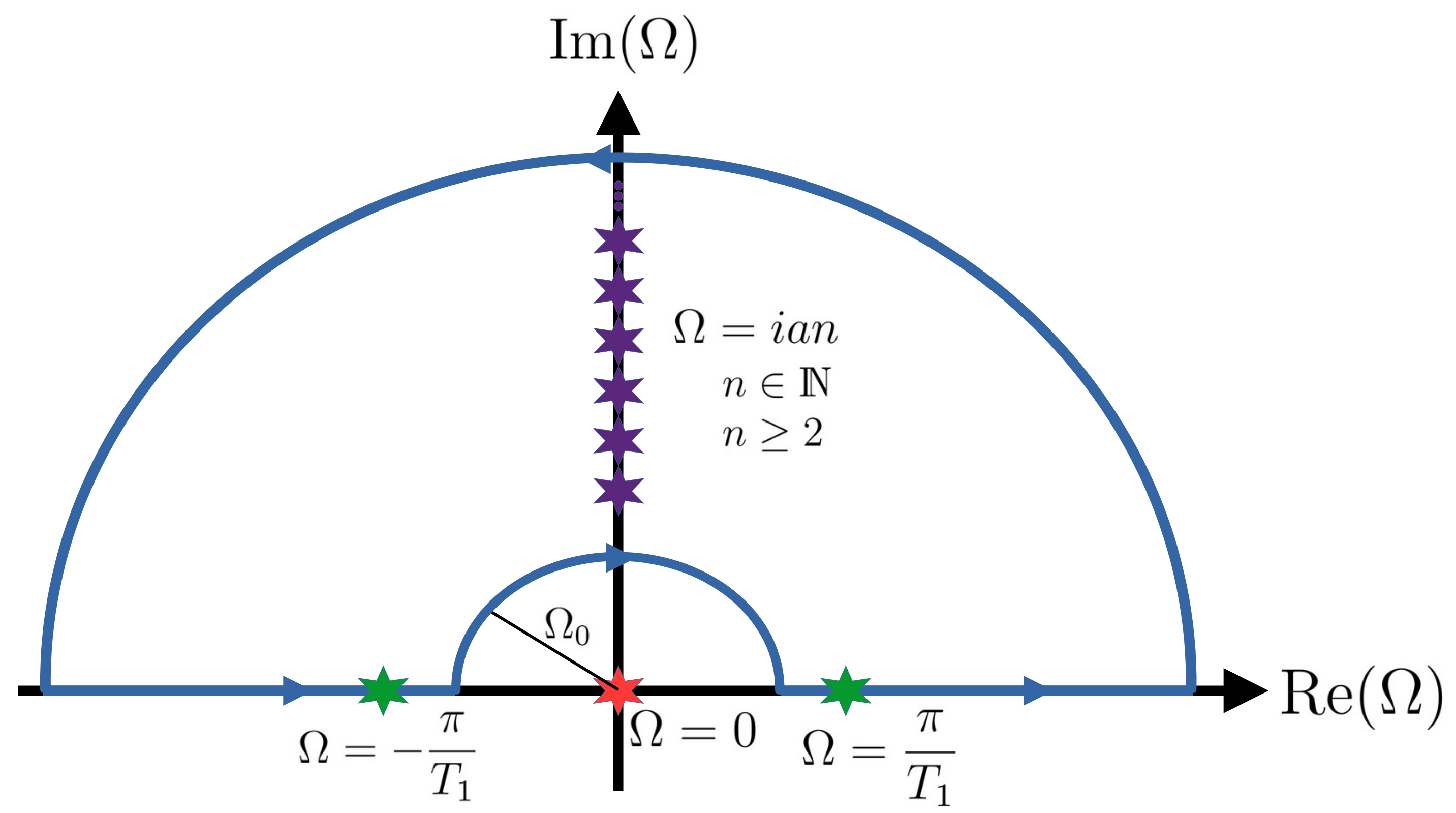}
  \caption{Sketch of the integration contour and poles in equation \eqref{eq:Dtwoterms} for the case without cut (left diagram) and with cut at $\Omega= \Omega_0$ (right diagram).}\label{fig:cont}
\end{figure*}

In both integrals the residue theorem can be applied. 
Given the form of the integrands, the contour can be closed using a semi-circle in the upper half plane, as the exponentials extended to the complex plane read
\begin{equation}
    e^{i \Omega \tau} \rightarrow e^{i |\Omega| \tau e^{i\phi}} = e^{i|\Omega| \tau \cos(\phi)} e^{-|\Omega| \tau \sin(\phi)}\,.
\end{equation}
As in all cases $\tau> 0$, this goes to zero for $|\Omega|\to\infty$ as long as $\sin(\phi) > 0$, which is precisely the case when closing the contour in the upper half plane. Additionally, for the term coming with $1$ instead of an exponential, the overall dependence on $\Omega$ is $\propto \frac{1}{|\Omega|^4}$ for large $|\Omega|$, hence also in that case the contribution of the semi-circle in the upper half plane with radius $|\Omega|\to\infty$ vanishes.
Therefore, the integral is completely determined by the contribution of the poles. A sketch of the integration contour and the poles can be found in the left diagram of figure~\ref{fig:cont}.

As shown in detail in Appendix \ref{ap:detcal}, the pole structure in equation \eqref{eq:Dtwoterms} is the following:
\begin{itemize}
    \item The first pole, which comes from $\mathds{D}_a$ in equation \eqref{eq:Dtwoterms}, is a pole of first order at $\Omega=0$ with      contribution
        \begin{equation}\label{eq:clashorresTP}
            \frac{C a}{2\pi} \left( \mathcal{T}+\frac{1}{2} \mathcal{T}_1\right)\,.
        \end{equation}
        This pole yields the linear behavior on $\mathcal{T}$ (with correct prefactor) obtained in the study of classical horizons discussed in section \ref{sec:ClasHor}.
    \item The term $\frac{\eta(\Omega,\mathcal{T},\mathcal{T}_1)}{\left(1 - \frac{\Omega^2 \mathcal{T}_1^2}{\pi^2}\right)^2}$, which is present in both $\mathds{D}_a$ and $\mathds{D}_b$, has two first order poles at $\Omega= \pm \frac{\pi}{\mathcal{T}_1}$. Their overall contribution combined from $\mathds{D}_a$ and $\mathds{D}_b$ is independent of $\mathcal{T}$ and reads
        \begin{equation}\label{contributionT1}
            \frac{C}{8} \left(\pi + \frac{\pi^3}{a^2 \mathcal{T}_1^2} \right) \coth\left( \frac{\pi^2}{a \mathcal{T}_1} \right)\,.
        \end{equation}
        As it is independent of $\mathcal{T}$, it contributes to the offset in the decoherence functional present due to the decoherence from the creation and destruction of the dipole. This is well visible in the fact that in the limit $\mathcal{T}_1\to 0$ this contribution diverges. This motivates the introduction of a non-zero time $\mathcal{T}_1>0$, as outlined above in sections \ref{sec:anofint} and \ref{sec:plotsdeco}
        (this term diverges also 
        in the ${\cal T}_1 \to \infty$ limit, but in this case
        the contribution is anyhow negligible w.r.t. that coming from $\cal T$ with our assumption $\mathcal{T}_1 \ll \mathcal{T}$).
        
    \item The remaining poles arise from the extension of the hyperbolic cotangent to the complex plane and are from $\mathds{D}_b$ in equation \eqref{eq:Dtwoterms}. These are first order poles at $\Omega =\pm i a n$ with $n\geq 2$, where only the poles at $\Omega = + i a n$ lie inside the integration contour. Their contribution is rather complicated and depends on $\mathcal{T}$ in a sum of combinations of exponentials, Lerch transcendents and Polygamma functions. However, as discussed in detail in Appendix~\ref{ap:detcal}, 
    for large $\mathcal{T}$ they contribute a finite, constant value independent of $\mathcal{T}$
    which adds to the saturation value from other terms.
    For small $\mathcal{T}$ they account for some features
    present at very small values of $\mathds{D}$. 
\end{itemize}
This shows that the linear behavior observed for classical horizons for large $\mathcal{T}$
arises solely from the pole at $\Omega=0$.

\subsubsection{Restricting the integration domain to the areas outside the cut region}\label{sec:quhres2}
Now, a small cut $\Omega_0$ is included into the decoherence functional which fulfills $\Omega_0 < 2a$ and $\Omega_0 < \frac{\pi}{\mathcal{T}_1}$. The first condition is fulfilled for all cases of interest mentioned in section \ref{sec:qualdisc}.
The second condition evaluates for $\Omega_0 <2a$ to
$\mathcal{T}_1 \, a < \frac{\pi}{2}$, 
and, as we will see in a moment, corresponds to leave unaltered
the contribution from $\mathcal{T}_1$, that is to keep
the term (\ref{contributionT1});
it also includes the cases of interest
for $\mathcal{T}_1$ small enough.
These requirements imply that the pole structure does not change apart from the pole at $\Omega=0$ compared to the case without cut discussed in the previous section. Hence the main change is the behavior of the integrand close to $\Omega=0$. 

The decoherence functional with the cut included in equation \eqref{eq:decfctC} can be rewritten as
\begin{align}
\mathds{D} = \frac{C}{4\pi a^2} \text{Re} \left(\int_{\Omega_0}^\infty + \int_{-\infty}^{-\Omega_0} \right) d\Omega\;&  \frac{a^2+\Omega^2}{\Omega}  \frac{\eta(\Omega,\mathcal{T},\mathcal{T}_1)}{\left(1-\frac{\Omega^2 \mathcal{T}_1^2}{\pi^2}\right)^2} \nonumber\\ &\hspace{0.4in}\cdot \coth\left(\frac{\pi\Omega}{a}\right)\,,
\end{align}
where $\eta(\Omega,\mathcal{T},\mathcal{T}_1)$ was defined in equation \eqref{eq:defeta}.
Similarly to the treatment of the case for the classical horizon, we use the same series expansion of the hyperbolic cotangent as in \eqref{eq:Dtwoterms} to obtain a similar split into $\mathds{D} = \mathds{D}_a^{\Omega_0} + \mathds{D}_b^{\Omega_0} $. To solve the integrations, we close the contour by a semi-circle in the upper half plane and add additionally a semi-circle in the upper half plane around the origin $\Omega=0$ with radius $\Omega_0$. A sketch of the integration contour and the poles can be found in the right diagram of figure \ref{fig:cont}.
Compared to the case in 
the previous subsection,
now instead of the pole at $\Omega=0$ we have to evaluate the contribution of the semi-circle with radius $\Omega_0$ around the origin $\Omega=0$. 

The detailed analysis is carried out in Appendix \ref{sec:Ap2}, here we present the main results.
\begin{itemize}
    \item $\mathds{D}_b^{\Omega_0}$ i.e., the contribution from the terms in the expansion of the hyperbolic cotangent where $\Omega=0$ was not a pole (that is the terms in the second line of \eqref{eq:Dtwoterms}, now with a cut included), yields a finite value independent of $\mathcal{T}$ which fulfills (see equation \eqref{eq:apbfdb} in the Appendix)
    \begin{equation}\label{D_b}
        \mathds{D}_b^{\Omega_0}\leq \frac{2C }{3} \frac{\frac{\Omega_0}{a}\left( 1+ \frac{\Omega_0^2}{a^2}\right)}{\left(1-\frac{\Omega_0^2 \mathcal{T}_1^2}{\pi^2}\right)^2}\,.
    \end{equation}
    When $\Omega_0 \mathcal{T}_1$  is way smaller
    than $\pi$,
    this gives
    $\mathds{D}_b^{\Omega_0} \ll 1 $. 
    \item For the term 
    in the expansion of the hyperbolic cotangent that leads to the pole at $\Omega=0$, that is $\mathds{D}_a^{\Omega_0}$  from the first line in equation \eqref{eq:Dtwoterms} but with cut, the semi-circle now circumvents the pole such that it is not any more on the integration contour. 
    Focusing on the regime $\Omega_0 \mathcal{T}_1$ small,
    the contribution that replaces the pole which led to a linear dependence on $\mathcal{T}$ now obtains a more complex form which can be found in equation \eqref{eq:newtermcutAp} in the Appendix.
    \end{itemize}

Therefore, when including a small cut, the main difference regarding the dependence of the decoherence functional on $\mathcal{T}$ comes from $\mathds{D}_a^{\Omega_0}$ which can be found in equation \eqref{eq:newtermcutAp}, which replaces the contribution of the pole at $\Omega=0$ in the case without the cut. 
While that contribution was given by
\eqref{eq:clashorresTP}, leading to a linear increase of decoherence with time $\mathcal{T}$
(and which can still be obtained from $\mathds{D}_a^{\Omega_0}$ in the limit of $\Omega_0 \to 0$), 
the situation changes drastically with a quantized horizon. In the limit of $\mathcal{T}\to\infty$, one now finds (using additionally $\Omega_0 \mathcal{T}_1 \ll 1$):
\begin{align}\label{eq:analresdeco}
    \mathds{D}_a^{\Omega_0}&\approx \frac{C a}{2 \pi} \left[ -\frac{\mathcal{T}_1}{2} + \frac{2}{\pi \Omega_0} - 2\frac{\Omega_0}{\pi a^2} \right] \nonumber\\ &\approx \frac{C a}{2 \pi} \left[\frac{2}{\pi \Omega_0} - 2\frac{\Omega_0}{\pi a^2} \right] \,,
\end{align}
which is constant and independent of $\mathcal{T}$, hence we arrive at a saturation of the decoherence no matter how small the actual value of the cut is. Compared to $\mathds{D}_b^{\Omega_0}$ in equation \eqref{D_b}, which can be bounded from above by $\sim C \frac{\Omega_0}{a}$, $\mathds{D}_a^{\Omega_0}$ is larger for $\mathcal{T}\to\infty$ for $\Omega_0 \lesssim \frac{a}{2}$ and of the same tiny order
or smaller for $\Omega_0$ larger
up to $\Omega_0 \sim a$.
\\ For small $\mathcal{T}$, there is a linear increase in $\mathds{D}_a^{\Omega_0}$, which can be seen by a Taylor expansion of the result \eqref{eq:newtermcutAp}. This Taylor expansion looses validity once the quadratic order reaches unity, that is once $\Omega_0 \mathcal{T} \sim 1$. In lowest order in $\Omega_0 \mathcal{T}_1$, the coefficient of the linear term in the expansion for small $\mathcal{T}$ is 
\begin{equation}
    \mathds{D}_a^{\Omega_0}\approx \frac{C a}{2\pi} \mathcal{T}\,,
\end{equation}
which coincides with the case without cut.
We see, the asymptotic value (\ref{eq:analresdeco})
plays the role of the
quantity $\mathds{D}_\text{sat}$ in the qualitative 
approach before in equation \eqref{eq:qualanalfirstDec},
$\mathds{D}_\text{sat} = 
\lim_{\mathcal{T}\to\infty}\mathds{D}_a^{\Omega_0}$,
where the case $\Omega_0 \ll a$ was discussed.

Going to $\Omega_0 \sim a$, the remaining part of the hyperbolic cotangent will become dominant, as the part discussed here in detail is only dominant for small argument $\frac{\pi \Omega_0}{a}$. 
The saturation is reached quicker given that it is approached after a time $\mathcal{T}\sim \frac{1}{\Omega_0}$ and the saturation value (\ref{eq:analresdeco})
becomes very small and
comparable or smaller than the (small) term (\ref{D_b}).
This analysis hence confirms the qualitative discussions from section \ref{sec:qualdisc}.

\section{Conclusions}

Building on previous results
\cite{Danielson:2022tdw, Danielson:2022sga, Gralla:2023oya, Wilson-Gerow:2024ljx, Danielson:2024yru},
in this work 
(which deepens and further develops \cite{FahnP})
the effects of the presence of a horizon on the coherence properties of a quantum system,
specifically an electric charge,
put in a spatial superposition near this horizon have been considered.
In \cite{Danielson:2022tdw, Danielson:2022sga, Gralla:2023oya, Wilson-Gerow:2024ljx, Danielson:2024yru},
for a horizon with a classical geometry, 
that is regarding the quantum fields sourced by the particle as propagating in a classical spacetime
background,
the result was found that the mere presence 
of the horizon induces decoherence on the particle's
system increasing linearly with the time the superposition
has been kept open. 
The attempt we have done here has been to consider
how this analysis gets modified if we take into account
some quantumness of the spacetime itself.

The element of quantumness we have chosen to use is
the existence of a minimal length, essentially
akin to the Planck length, even if not necessarily exactly so.
This is something foreseen in most full fledged quantum
approaches to gravity, at least at an effective level.
Previous results \cite{QuantumMetricNull, KriPer, DICE} have shown that this implies the existence of a quantum of area change in any process
involving exchange of energy between the horizon and its surroundings, and reconnects then to quantum black holes for which the quantum aspects of spacetime are
introduced as black hole area quantization \cite{Bekenstein, Maggiore, Hod, Barbero2}.

The classical spacetime results have been obtained both
in a global description,
in which the properties of the quantum fields sourced 
by the particle are considered on the whole (curved) spacetime \cite{Danielson:2022tdw, Danielson:2022sga, Gralla:2023oya}, and in a local description in which the
fields are considered at the location
where the particle is put in superposition and then recombined
\cite{Wilson-Gerow:2024ljx, Danielson:2024yru}.
One would like the two perspectives to produce similar
results, and this has turned out indeed to be the case
in suitable limits (of long time persistent effects),
with the main contribution to decoherence coming
from soft modes of the field piercing the horizon,
this mechanism also underlying the enhancement of radiation reaction
(responsible of the linear increase of decoherence in time
in the local description)
as perceived by the accelerating 
quantum system
\cite{Danielson:2022tdw, Danielson:2022sga, Gralla:2023oya, Wilson-Gerow:2024ljx, Danielson:2024yru}.

In the present work, 
the existence of a quantum of area 
has been taken to imply the existence of a cut 
on frequency for the modes of the field which can
pierce the horizon:
Only modes with frequency above some threshold frequency
(determined by the parameters characterizing the horizon)
can be absorbed, by energy conservation.
This is something similar to what has already been explored in
the context of possible effects of existence of a quantum of area on tidal heating in the inspiral phase, or on ringing modes in the ringdown, of the coalescence of black hole binaries and their gravitational echos \cite{Cardoso:2019, Agullo:2021, Chakravarti:2021jbv, Chakravarti:2021clm, KriPer, DICE} or in the context of quantized angular momenta for rotating black holes \cite{Emparan:2025qqf}. 

Focusing on the case of a Schwarzschild black hole, 
what we find is that the presence of this frequency cut non-negligibly affects, generically, the decoherence.
The main effect is to turn the increase in decoherence, which goes linear in the time the superposition has been kept open,
into a linear increase only up to a certain critical time,
above which the decoherence stops increasing and approaches a constant asymptotic value.
Point is that this asymptotic value for decoherence
turns out to be quite small, even for quanta of horizon area really small as compared to those reported in the literature in the years.
This means that for the latter the maximum decoherence
we can reach turns out to be very small, 
something $\sim 10^{-4}-10^{-5}$ at the most
(choosing the electric charge to be the electron charge $e$),
then somehow of very limited effect.
In other words,
if the quantum properties of spacetime are implemented
as described here, namely in terms of existence of
a limit length or of existence of a quantum of 
horizon area,
we have that quantum properties of the geometry of
the horizon actually suppress the decohering action
from the latter to a negligible level, 
unless the quanta of area are way smaller than the selected values proposed up to now.

The limiting values of decoherence,
e.g. for the value of a quantum of area 
$8\pi l_p^2$ \cite{Bekenstein, Maggiore}
(which gives the asymptotic angular frequency cut 
$\omega_0 = \kappa$, with $\kappa$ the surface gravity
of the horizon)
or for the value $4 l_p^2 \ln 3$
\cite{Hod, Barbero2}
(which gives  $\omega_0 = \frac{\ln 3}{2 \pi} \kappa$),
turn out to be so small to be comparable to the
decoherence due to modes reaching to null infinity
(considered e.g. in   \cite{Belenchia:2018szb,Danielson:2022sga,Gralla:2023oya}),
that is what one gets already in absence of a horizon,
and to the effects from the transients
in the separation and recombination procedures for the particle under consideration,
however small we want them.
The latter aspect forced us indeed to consider
an explicit model for the transients.
Coming to the view of black holes
as thermal black quantum systems at finite temperature \cite{Biggs:2024dgp},
we see that for these values of the quantum of 
area 
there is a challenge to the possibility to treat 
the energy exchanges in the continuum limit
when in the low energy limit.

A last comment is about the fate, in view of these results,
of the gedankenexperiment from which we started
our investigation
\cite{Mari:2015qva, Belenchia:2018szb, WalG, WaldH}.
We saw that, in case the mass $B$ is beyond the horizon,
$A$ can recombine in a very long time still having
no information from $B$,
and this challenges the coexistence of causality and complementarity unless the horizon itself, with its mere presence, induces decoherence on $A$.
Now, our analysis shows that,
if certain quantum aspects of the horizon are taken
into account, the horizon may no longer be able to
induce appreciable decoherence,
this apparently bringing again to a potential clash
between causality and complementarity.

Our results do suggest however also a way out:
If the soft modes of the field cannot pierce the quantum horizon,
having thus the latter sort of reflectivity at low 
frequencies, 
no relevant information about $A$ can reach $B$,
and $B$ can not discriminate the path of $A$,
this then solving again the paradox.
In other words: a) in case the geometry of the horizon can be treated as classical,
soft radiation pierces the horizon and $B$ can do which path; $A$ knows nothing from $B$, but the horizon induces 
anyway decoherence on $A$, and this solves the paradox;
b) in case the horizon geometry 
has to be treated as quantum, we might even have 
only negligible decoherence on $A$, but then
soft radiation would not pierce the horizon and 
$B$ would be unable to do which-path.
We see then the paradox appears to be solved, either way.
This case shows hence some similarities with a maximally rotating Kerr black hole with classical geometry, which prevents the entire field modes from entering the horizon \cite{Gralla:2023oya}.

We expect that our analysis can, similarly as in \cite{Danielson:2022tdw,Gralla:2023oya,Danielson:2022sga,Wilson-Gerow:2024ljx,Danielson:2024yru}, also be applied to a linearized gravitational field instead of a photon field. 
In this formulation,
when considered in the local description above,
it would provide an application of the theory of open quantum systems with quantized gravity as environment,
something studied in recent years for instance in the context of cosmology (see e.g. \cite{Boyanovsky:2015tba,boyanovsky2015effective,hollowood2017decoherence,martin2018observational,colas2022benchmarking,brahma2024time}) and gravitational waves (see e.g. \cite{Blencowe:2012mp,anastopoulos2013master,oniga2016quantum,bassi2017gravitational,lagouvardos2021gravitational,fahn2023gravitationally,Domi:2024ypm,fgSoon,cho2025non}).

\acknowledgments 
The authors would like to thank Roberto Casadio for fruitful discussions and suggestions. 
Their research is partially supported
by the INFN grant FLAG.
MJF also would like to acknowledge the
contribution of the COST action
CA23115, 
and both authors the COST action
CA23130.

\onecolumngrid
\vspace{0.4in}
\hrule
\appendix

\section{More details on the chosen dipole moment}\label{sec_App0}
In section \ref{sec:dipmom}, a specific choice of a dipole moment was introduced in order to extract some more details from the decoherence functional. In this Appendix, we discuss this dipole moment further and determine the decoherence the corresponding dipole experiences due to the presence of the classical horizon discussed in the main text. \\
The definition of the dipole moment $\epsilon_1(\tau)$ in equation \eqref{eq:dpmom} allows to evaluate its Fourier transform as
\begin{align}
    \tilde{\epsilon}_1(\Omega) =& \int_\mathds{R} d\tau\; e^{-i\Omega\tau} \epsilon_1(\tau)
    = \frac{2 d}{\Omega} \frac{1}{1 - \frac{\Omega^2 \mathcal{T}_1^2}{\pi^2}} \sin\left( \frac{\mathcal{T}+\mathcal{T}_1}{2}\Omega \right) \cos\left( \frac{\mathcal{T}_1}{2}\Omega \right) \,.
\end{align}
For large $\Omega$, this function goes to zero as $\Omega^{-3}$, such that the decoherence functional $\mathds{D}$ in \eqref{eq:defdecfctft} is UV finite. If one sends $\mathcal{T}_1 \rightarrow 0$ at this point or equally assumes instantaneous creation/destruction of the dipole, $\tilde{\epsilon}_1(\Omega)$ behaves as $\Omega^{-1}$ for $\Omega\to \infty$ and hence the decoherence functional would diverge quadratically for large $\Omega$. Also the choice of for instance a linear behavior of $\epsilon_1(\tau)$ in the time intervals $\left[ -\frac{\mathcal{T}}{2}-\mathcal{T}_1, -\frac{\mathcal{T}}{2}\right]$ and $\left[ \frac{\mathcal{T}}{2}, \frac{\mathcal{T}}{2}+\mathcal{T}_1\right]$
would give $\tilde{\epsilon}_1(\Omega) \to \Omega^{-2}$, which would imply a logarithmic UV divergence of $\mathds{D}$.

Due to this, we work with $\epsilon_1(\tau)$ as defined above, where the presence of $\mathcal{T}_1$ acts as a UV regulator. This culminates in the following decoherence function:
\begin{align}\label{eq:decfctap}
    \mathds{D} &= \frac{2C}{\pi a^2} \int_{0}^\infty d\Omega\;  \frac{a^2+\Omega^2}{\Omega}  \frac{\cos^2\left(\frac{\mathcal{T}_1}{2}\Omega\right)}{\left(1-\frac{\Omega^2 \mathcal{T}_1^2}{\pi^2}\right)^2}\sin^2\left(\frac{\mathcal{T}+\mathcal{T}_1}{2}\Omega\right) \coth\left(\frac{\pi\Omega}{a}\right)\,.
\end{align}
Aside from the UV finiteness, for $\Omega\ll 1$ we have that $\sin^2\left(\frac{\mathcal{T}+\mathcal{T}_1}{2}\Omega\right) \sim \Omega^2$ precisely cancels $\frac{1}{\Omega}\coth\left(\frac{\pi \Omega}{a}\right) \sim \frac{1}{\Omega^2}$, thus yielding also a finite value at $\Omega=0$. At $\Omega = \frac{\pi}{\mathcal{T}_1}$, the cosine renders the integrand finite albeit the presence of the denominator which is zero at that point. To preserve the spirit of the initial gedankenexperiment, we focus on the case where $\mathcal{T} \gg \mathcal{T}_1$.

Let us continue by showing that, by some arguments brought up in \cite{Gralla:2023oya}, we can obtain an approximation for the decoherence functional \eqref{eq:decfctap} which is equal to the last line of equation \eqref{eq:Dfctapro1} in the main text in the limit of large $\mathcal{T}$. 
The derivation of that equation, 
along the lines of  \cite{Wilson-Gerow:2024ljx}, 
heavily relies on the fact that the integration over $\Omega$ includes very small frequencies, which are the reason for the decoherence process linear in $\mathcal{T}$. In \cite{Gralla:2023oya}, the treatment is slightly more general and in this section we apply a reasoning very similar to the one in \cite{Gralla:2023oya} to our specific decoherence functional in equation \eqref{eq:decfctap} in order to get a better insight into the behavior of the integrand away from $\Omega=0$. 
The entire integration is split into two parts, where the first one ranges over the interval $[0,\Omega_s]$ with a split frequency $\Omega_s$ and the remaining interval $[\Omega_s,\infty]$. The split frequency is assumed to fulfill
\begin{equation}\label{eq:splitfrequ}
    \Omega_s \mathcal{T} \gg 1 \hspace{0.2in} \text{and} \hspace{0.2in} \Omega_s \mathcal{T}_1\ll 1\,.
\end{equation}
With this, one can approximate the decoherence functional $\mathds{D}$ in equation \eqref{eq:decfctap} as
\begin{align}
     \mathds{D} \approx& \frac{2C}{\pi a^2} \int_{0}^{\Omega_s} d\Omega\;  \frac{a^2+\Omega^2}{\Omega}  \sin^2\left(\frac{\mathcal{T}}{2}\Omega\right)   \coth\left(\frac{\pi\Omega}{a}\right)\nonumber\\
     &+\frac{2 C}{\pi a^2} \int_{\Omega_s}^\infty d\Omega\;  \frac{a^2+\Omega^2}{\Omega}  \frac{\cos^2\left(\frac{\mathcal{T}_1}{2}\Omega\right)}{\left(1-\frac{\Omega^2 \mathcal{T}_1^2}{\pi^2}\right)^2}  \sin^2\left(\frac{\mathcal{T}+\mathcal{T}_1}{2}\Omega\right)  \coth\left(\frac{\pi\Omega}{a}\right)\,.
\end{align}
In order to satisfy the conditions in \eqref{eq:splitfrequ} for any choice of $\mathcal{T}$ and $\mathcal{T}_1$, the split frequency can depend on $\mathcal{T}$ and $\mathcal{T}_1$, one possible choice with correct dimensions being $\Omega_s = \frac{1}{\sqrt{\mathcal{T} \mathcal{T}_1}}$. For fixed $\mathcal{T}_1$, waiting long enough in $\mathcal{T}$ will lead into a regime where $\frac{\Omega_s}{a}\ll 1$, which allows to approximate the hyperbolic cotangens in the first integral as 
\begin{equation}
    \coth\left(\frac{\pi\Omega}{a}\right) \approx \frac{a}{\pi\Omega}\,.
\end{equation}
In the second integral, one can use the condition $\Omega_s \mathcal{T}\gg 1$ to average over the fast rotating terms yielding a factor of $\frac{1}{2}$ and a correction term being of the order $\frac{\mathcal{T}_1}{\mathcal{T}}$, which in summary leads to
\begin{align}
     \mathds{D} \approx& \underbrace{\frac{2 C}{\pi^2 a} \int_{0}^{\Omega_s} d\Omega\;  \frac{a^2}{\Omega^2}  \sin^2\left(\frac{\mathcal{T}}{2}\Omega\right)}_{=: \mathds{D}_1}  +\underbrace{\frac{C}{\pi a^2} \int_{\Omega_s}^\infty d\Omega\;  \frac{a^2+\Omega^2}{\Omega}  \frac{\cos^2\left(\frac{\mathcal{T}_1}{2}\Omega\right)}{\left(1-\frac{\Omega^2 \mathcal{T}_1^2}{\pi^2}\right)^2}   \coth\left(\frac{\pi\Omega}{a}\right)}_{=: \mathds{D}_2}\,.
\end{align}
Substituting in the first integral $\Omega \rightarrow \Omega \mathcal{T} =: \lambda$ gives
\begin{equation}\label{eq:solgral1}
    \mathds{D}_1 =  \frac{2C}{\pi^2 a} \int_{0}^{\Omega_s \mathcal{T}} d\lambda\;  \frac{a^2 \mathcal{T}}{\lambda^2}  \sin^2\left(\frac{\lambda}{2}\right) \approx \frac{C a}{2 \pi} \mathcal{T}\,,
\end{equation}
where in the last step we sent $\Omega_s \mathcal{T} \to \infty$. The second part $\mathds{D}_2$ does not directly depend on $\mathcal{T}$ any more, only via the split frequency $\Omega_s$. The argument in \cite{Gralla:2023oya} is now the following: The result in equation \eqref{eq:solgral1} for $\mathds{D}_1$ is the leading order in an expansion for large $\mathcal{T}$ and is independent of the split frequency. As this split frequency was introduced artificially, the entire integral in $\mathds{D}$ should not depend on it, either. Hence the dependence has to cancel in each order of the $\mathcal{T}$ expansion. As the leading order from the expansion of $\mathds{D}_1$ yields a linear dependence on $\mathcal{T}$ without dependence on $\Omega_s$, the contribution of $\mathds{D}_2$ can only affect higher orders in the expansion, therefore for large $\mathcal{T}$ one obtains
\begin{equation}
    \mathds{D} \approx  \frac{C a}{2 \pi} \mathcal{T}\,,
\end{equation}
which is in accordance with equation \eqref{eq:Dfctapro1} and thus with the results from \cite{Danielson:2022tdw,Danielson:2022sga,Gralla:2023oya,Wilson-Gerow:2024ljx,Danielson:2024yru}.

\section{Detailed analytical calculation for horizons with classical and quantum geometries}\label{ap:detcal}
In this Appendix, we present the detailed discussion and calculation of the decoherence caused by a (quantized) horizon with a small frequency cut using residue theorem techniques, which was summarized in section \ref{sec:quhres1} in the main text. In the first part, we consider the entire real line as integration domain, that is we treat the cut as not existent and hence expect to restore the result obtained for a classical horizon. In the second part, we then introduce a small cut frequency and investigate the changes of the decoherence functional due to this.

\subsection{Horizon with classical geometry}\label{sec:Ap1}
We start from the decoherence functional in the form of equation \eqref{eq:Dtwoterms}:
\begin{align}\label{eq:DtwotermsAp}
    \mathds{D} =& \underbrace{\frac{C}{4\pi^2 a} \text{Re} \int_{-\infty}^\infty d\Omega\;  \frac{a^2+\Omega^2}{\Omega^2}  \frac{1}{\left(1-\frac{\Omega^2 \mathcal{T}_1^2}{\pi^2}\right)^2} \left[ 1 - e^{i\Omega(\mathcal{T}+\mathcal{T}_1)} + e^{i\Omega \mathcal{T}_1} - \frac{1}{2} e^{i \Omega \mathcal{T}} -\frac{1}{2} e^{i \Omega (\mathcal{T}+2 \mathcal{T}_1)}\right]}_{=:\mathds{D}_a} \nonumber\\ &+ \underbrace{\frac{C}{2\pi^2 a} \text{Re} \sum_{n=1}^\infty \int_{-\infty}^\infty d\Omega\;  \frac{a^2+\Omega^2}{a^2 n^2 + \Omega^2}  \frac{1}{\left(1-\frac{\Omega^2 \mathcal{T}_1^2}{\pi^2}\right)^2} \left[ 1 - e^{i\Omega(\mathcal{T}+\mathcal{T}_1)} + e^{i\Omega \mathcal{T}_1} - \frac{1}{2} e^{i \Omega \mathcal{T}} -\frac{1}{2} e^{i \Omega (\mathcal{T}+2 \mathcal{T}_1)}\right]}_{=:\mathds{D}_b}\,.
\end{align}
To solve this integration, we go into the complex plane and consider a contour which we close with a semi circle in the upper half plane of infinite radius. As discussed in the main text in section \ref{sec:analp1}, its contribution vanishes. What remains is to analyze the pole structure and their contributions for each individual term. The resulting pole structure is visualized in the left diagram of figure \ref{fig:cont}.\\~\\
\textbf{Evaluation of $\mathds{D}_a$}\\
The integrand in $\mathds{D}_a$ in equation \eqref{eq:DtwotermsAp} has two points that might be problematic and yield poles.
\begin{itemize}
    \item At $\Omega\to 0$, both numerator and denominator give zero, where the denominator goes to zero as $\frac{1}{\Omega^2}$. Applying L'H{$\hat {\rm o}$}pital's rule once gives a nonzero value for the numerator, so the integrand has a pole of first order (simple pole) at $\Omega=0$.
    Its contribution is, given that the pole lies on the contour and hence it contributes $\frac{1}{2}$ of its usual contribution (using the indentation lemma\footnote{See \cite{indentationLemma}; alternatively one can also consider the contribution of a semi-circle around the pole with vanishing radius and obtains the same result, which is similar to what is done in part \ref{sec:Ap2} of this Appendix when including the cut.}):
    \begin{align}
        &\frac{C}{4 \pi^2 a}\text{Re} \left[\pi i \frac{a^2+\Omega^2}{\Omega} \frac{1}{\left(1-\frac{\Omega^2 \mathcal{T}_1^2}{\pi^2}\right)^2} \left( 1 - e^{i\Omega(\mathcal{T}+\mathcal{T}_1)} + e^{i\Omega \mathcal{T}_1} - \frac{1}{2} e^{i \Omega \mathcal{T}} -\frac{1}{2} e^{i \Omega (\mathcal{T}+2 \mathcal{T}_1)}\right) \right]_{\Omega=0}\nonumber\\
        &=\frac{C}{4 \pi^2 a}\text{Re} \left[a^2 \pi i (-i (2\mathcal{T}+\mathcal{T}_1)) \right] = \frac{C a}{2 \pi} \left(\mathcal{T}+\frac{1}{2}\mathcal{T}_1\right)\,,
    \end{align}
    where we made use of L'H{$\hat {\rm o}$}pital's rule from above for evaluating the sum of the exponentials over $\Omega$ at $\Omega=0$. This pole leads to the linear dependence in $\mathcal{T}$ (with the correct prefactor) obtained by the previous works \cite{Danielson:2022tdw,Danielson:2022sga,Gralla:2023oya,Wilson-Gerow:2024ljx,Danielson:2024yru} and discussed in section \ref{sec:ClasHor}.
    \item The second problematic point in $\mathds{D}_a$ in equation \eqref{eq:DtwotermsAp} is at $\Omega \to \pm \frac{\pi}{\mathcal{T}_1}$. To discuss this further, we rewrite
    \begin{equation}
        \frac{1}{\left(1-\frac{\Omega^2 \mathcal{T}_1^2}{\pi^2}\right)^2} = \frac{\pi^4}{\mathcal{T}_1^4} \frac{1}{\left(\Omega + \frac{\pi}{\mathcal{T}_1}\right)^2 \left(\Omega - \frac{\pi}{\mathcal{T}_1}\right)^2}\,.
    \end{equation}
    Similarly to the case above, also the sum of exponentials gives zero at $\Omega = \pm \frac{\pi}{T_1}$. L'H{$\hat {\rm o}$}pital's rule shows that there is a pole of first order.    
    The contribution of this pole (again making use of the indentation lemma, as the pole lies on the integration contour) is
    \begin{align}
        \frac{C}{4\pi^2 a} \text{Re} \left[ \pi i \frac{a^2 + \frac{\pi^2}{\mathcal{T}_1^2}}{\frac{\pi^2}{\mathcal{T}_1^2}} \frac{\pi^4}{\mathcal{T}_1^4} \frac{1}{4 \frac{\pi^2}{\mathcal{T}_1^2}} (-i \mathcal{T}_1)\right] = \frac{C}{16 \pi a} \left(a^2 + \frac{\pi^2}{\mathcal{T}_1^2} \right) \mathcal{T}_1\,.
    \end{align}
    independently of the sign of the pole, so their overall contribution is
    \begin{equation}\label{eq:appolel1rl}
         \frac{C}{8 \pi a} \left(a^2 + \frac{\pi^2}{\mathcal{T}_1^2} \right) \mathcal{T}_1\,,
    \end{equation}
    which is independent of $\mathcal{T}$.
\end{itemize}
This concludes the discussion of the poles contributing to $\mathds{D}_a$.\\~\\
\textbf{Evaluation of $\mathds{D}_b$}\\
In the integral for $\mathds{D}_b$ in equation \eqref{eq:DtwotermsAp}, there are again two critical points for each $n$.
\begin{itemize}
    \item At the point $\Omega\to \pm\frac{\pi}{\mathcal{T}_1}$ there is a pole of first order on the contour, the discussion follows along the lines of the one for the term in $\mathds{D}_a$ above. The contribution of both these poles together to $\mathds{D}_b$ is
    \begin{align}
         - \frac{C}{8 \pi a} \left(a^2 + \frac{\pi^2}{\mathcal{T}_1^2} \right) \mathcal{T}_1 + \frac{C}{8} \left(\pi + \frac{\pi^3}{a^2 \mathcal{T}_1^2} \right) \coth\left( \frac{\pi^2}{a \mathcal{T}_1} \right) \,.
    \end{align}
    
    This does indeed cancel the contribution from the same pole in $\mathds{D}_a$ in equation \eqref{eq:appolel1rl} and incorporates it into the full hyperbolic cotangent. Additionally, one can see that the smaller the preparation time $\mathcal{T}_1$ gets, the larger becomes the decoherence. This contribution is independent of $\mathcal{T}$, hence it gives an offset in the decoherence present due to the preparation (and destruction) of the superposition.
    \item The last critical point in $\mathds{D}_b$ is at $\Omega = \pm i a n$. As we close the contour in the upper half plane, only the points at $\Omega = + i a n$ lie inside the integration area. Given the numerator $a^2+\Omega^2$, we see that for $n=1$ it cancels precisely the denominator, thus no pole is present here. Hence the poles arise for $n\geq 2$ as poles of first order. As the contour circumvents them counterclockwise, their contribution is
    \begin{align}
        &\frac{C}{2\pi^2 a} \text{Re} \sum_{n=2}^\infty \frac{a^2(1-n^2)}{2i a n} \frac{2\pi i}{\left(1+\frac{a^2 \mathcal{T}_1^2}{\pi^2} n^2\right)^2} \left[ 1 -   e^{-an(\mathcal{T}+\mathcal{T}_1)} + e^{-an \mathcal{T}_1} - \frac{1}{2} e^{-an \mathcal{T}} -\frac{1}{2} e^{-an (\mathcal{T}+2 \mathcal{T}_1)}\right]\nonumber\\
        &= \frac{C \pi^3 }{2 a^4 \mathcal{T}_1^4 } \sum_{n=2}^\infty \frac{1-n^2}{n} \frac{1}{\left(\frac{\pi^2}{a^2 \mathcal{T}_1^2} + n^2 \right)^2}\left[ 1 -   e^{-an(\mathcal{T}+\mathcal{T}_1)} + e^{-an \mathcal{T}_1} - \frac{1}{2} e^{-an \mathcal{T}} -\frac{1}{2} e^{-an (\mathcal{T}+2 \mathcal{T}_1)}\right]\,.
    \end{align}
    To evaluate this sum, we split it into two different parts.
    \begin{itemize}
    \item As first part, we consider
    \begin{align}
        &\frac{C \pi^3 }{2 a^4 \mathcal{T}_1^4 } \sum_{n=l}^\infty \frac{-n^2}{n} \frac{1}{n^4} \left[1-   e^{-an(\mathcal{T}+\mathcal{T}_1)} + e^{-an \mathcal{T}_1} - \frac{1}{2} e^{-an \mathcal{T}} -\frac{1}{2} e^{-an (\mathcal{T}+2 \mathcal{T}_1)}\right]\nonumber\\
        &=\frac{C \pi^3 }{4 a^4 \mathcal{T}_1^4 }  \Big(e^{- a l \mathcal{T}} \text{LP}\left[e^{-a \mathcal{T}},3,l\right]-2 e^{-a l \mathcal{T}_1} \text{LP}\left[e^{-a
\mathcal{T}_1},3,l\right] + \text{PolyGamma}[2,l]\nonumber\\ &\hspace{0.9in}+2 e^{-a l(\mathcal{T}+\mathcal{T}_1)} \text{LP}\left[e^{-a (\mathcal{T}+\mathcal{T}_1)},3,l\right]+ e^{-a l (\mathcal{T}+2 \mathcal{T}_1)}\text{LP}\left[e^{-a (\mathcal{T}+2 \mathcal{T}_1)},3,l\right]\Big)\,,
    \end{align}
    which is a good approximation for $l^2\gg 1$ and $l \gg \frac{\pi}{a \mathcal{T}_1}$. The Lerch transcendent is defined as
    \begin{equation}
        \text{LP}(x,y,z) = \sum_{n=0}^\infty \frac{x^n}{(n+z)^y} \,.
    \end{equation}
    It converges here as either $|x|<1$ and $z>0$ or $|x|=1$ and $y>1$. This allows us to take the limit $\mathcal{T}\to \infty$, where the contribution to the decoherence functional is:
    \begin{equation}
        \longrightarrow \frac{C \pi^3 }{4  a^4 \mathcal{T}_1^4 }\left(\text{PolyGamma}[2,l] -2 e^{-a l \mathcal{T}_1} \text{LP}\left[e^{-a \mathcal{T}_1},3,l\right]\right)\,.
    \end{equation}
    \item The second part contains the terms of the sum between a lowest value $m$ (which is $2$ here) and an upper value $(l-1)$, after which the approximation made above is reasonable:
    \begin{align}
        \frac{C \pi^3 }{2 a^4 \mathcal{T}_1^4 } \sum_{n=m}^{l-1} \frac{1-n^2}{n} \frac{1}{\left(\frac{\pi^2}{a^2 \mathcal{T}_1^2} + n^2 \right)^2}\left[1 - e^{-an(\mathcal{T}+\mathcal{T}_1)} + e^{-an \mathcal{T}_1} - \frac{1}{2} e^{-an \mathcal{T}} -\frac{1}{2} e^{-an (\mathcal{T}+2 \mathcal{T}_1)}\right]\,.
    \end{align}
    The sum can be computed and yields in the limit $\mathcal{T}\to\infty$ a sum of different products of exponentials and Lerch transcendents or Polygamma functions, which is finite:
    \begin{align}
        \frac{C}{8\pi a^4 \mathcal{T}_1^4 }\Bigg\{& -4 a^4 \mathcal{T}_1^4e^{-a l \mathcal{T}_1}  \text{LP}\left[e^{-a \mathcal{T}_1},1,l\right] +4 a^4 \mathcal{T}_1^4 e^{-a
m \mathcal{T}_1}  \text{LP}\left[e^{-a \mathcal{T}_1},1,m\right]\nonumber\\ &+2 a^4 \mathcal{T}_1^4 e^{-a l \mathcal{T}_1} \left(
\text{LP}\left[e^{-a \mathcal{T}_1},1,l-\frac{i \pi }{a \mathcal{T}_1}\right] +\text{LP}\left[e^{-a \mathcal{T}_1},1,l+\frac{i
\pi }{a \mathcal{T}_1}\right]\right)\nonumber\\ &-2 a^4 \mathcal{T}_1^4 e^{-a m \mathcal{T}_1} \left( \text{LP}\left[e^{-a
\mathcal{T}_1},1,m-\frac{i \pi }{a \mathcal{T}_1}\right] + \text{LP}\left[e^{-a \mathcal{T}_1},1,m+\frac{i \pi }{a
\mathcal{T}_1}\right] \right)\nonumber\\ &-i (a \mathcal{T}_1 \pi^3 + a^3 \mathcal{T}_1^3 \pi) e^{-a l \mathcal{T}_1}  \left(\text{LP}\left[e^{-a \mathcal{T}_1},2,l-\frac{i \pi }{a \mathcal{T}_1}\right]  -\text{LP}\left[e^{-a \mathcal{T}_1},2,l+\frac{i \pi }{a \mathcal{T}_1}\right] \right)\nonumber\\ 
&+i (a \mathcal{T}_1 \pi^3 + a^3 \mathcal{T}_1^3 \pi) e^{-a m \mathcal{T}_1} \left(\text{LP}\left[e^{-a \mathcal{T}_1},2,m-\frac{i \pi }{a \mathcal{T}_1}\right] -\text{LP}\left[e^{-a \mathcal{T}_1},2,m+\frac{i \pi }{a \mathcal{T}_1}\right]   \right)\nonumber\\  &+4 a^4 \mathcal{T}_1^4 \text{PolyGamma}[0,l]-4
a^4 \mathcal{T}_1^4 \text{PolyGamma}[0,m]\nonumber\\ &-2 a^4 \mathcal{T}_1^4 \left( \text{PolyGamma}\left[0,l-\frac{i \pi }{a \mathcal{T}_1}\right] + \text{PolyGamma}\left[0,l+\frac{i \pi }{a \mathcal{T}_1}\right]  \right) \nonumber\\&+2 a^4 \mathcal{T}_1^4 \left(\text{PolyGamma}\left[0,m-\frac{i
\pi }{a \mathcal{T}_1}\right] + \text{PolyGamma}\left[0,m+\frac{i
\pi }{a \mathcal{T}_1}\right]\right)\nonumber\\ &-i (a \mathcal{T}_1 \pi^3+a^3 \mathcal{T}_1^3 \pi) \left(\text{PolyGamma}\left[1,l-\frac{i \pi }{a \mathcal{T}_1}\right]-\text{PolyGamma}\left[1,l+\frac{i
\pi }{a \mathcal{T}_1}\right] \right)\nonumber\\ &+i (a \mathcal{T}_1 \pi^3+a^3 \mathcal{T}_1^3 \pi) \left(\text{PolyGamma}\left[1,m-\frac{i \pi }{a \mathcal{T}_1}\right]-\text{PolyGamma}\left[1,m+\frac{i
\pi }{a \mathcal{T}_1}\right] \right)\Bigg\}\,.
    \end{align}
    \end{itemize}
\end{itemize}
This shows all together that the linear dependence of the decoherence functional on $\mathcal{T}$ for large $\mathcal{T}$ arises solely from $\mathds{D}_a$, where it originates in the pole at $\Omega=0$. The remaining poles contribute a finite decoherence functional in the limit of $\mathcal{T}\to\infty$.

\subsection{Horizon with quantum geometry}\label{sec:Ap2}
As discussed in the main text in section \ref{sec:quhres2}, we now include a very small cut $\Omega_0$ (in the sense that $\Omega_0 < \frac{\pi}{\mathcal{T}_1}$ and $\Omega_0 < 2a$) into the decoherence functional. This has as effect that the pole structure does not change apart from the pole at $\Omega=0$ and that the same poles as above contribute. The main change is now the behavior close to $\Omega=0$. The decoherence functional with the cut included reads
\begin{equation}
    \mathds{D} = \frac{2C}{\pi a^2} \int_{\Omega_0}^\infty d\Omega\;  \frac{a^2+\Omega^2}{\Omega}  \frac{1}{\left(1-\frac{\Omega^2 \mathcal{T}_1^2}{\pi^2}\right)^2} \sin^2\left(\frac{\mathcal{T}+\mathcal{T}_1}{2}\Omega\right) \cos^2\left(\frac{\mathcal{T}_1}{2}\Omega\right)  \coth\left(\frac{\pi\Omega}{a}\right)\,,
\end{equation}
which we rewrite as 
\begin{align}\label{eq:DtwotermsCAp}
     \mathds{D} &= \frac{C}{4\pi a^2} \text{Re} \left(\int_{\Omega_0}^\infty + \int_{-\infty}^{-\Omega_0} \right) d\Omega\;  \frac{a^2+\Omega^2}{\Omega}  \frac{1}{\left(1-\frac{\Omega^2 \mathcal{T}_1^2}{\pi^2}\right)^2} \left[ 1 - e^{i\Omega(\mathcal{T}+\mathcal{T}_1)} + e^{i\Omega \mathcal{T}_1} - \frac{1}{2} e^{i \Omega \mathcal{T}} -\frac{1}{2} e^{i \Omega (\mathcal{T}+2 \mathcal{T}_1)}\right] \coth\left(\frac{\pi\Omega}{a}\right)\,.
\end{align}
Similarly to the case above, we can now use the series expansion of the hyperbolic cotangent and close the contour by a semi-circle in the upper half plane and a semi-circle around the origin $\Omega=0$ with radius $\Omega_0$. This second semi-circle is new compared to the case without the cut. A sketch of the contour and the poles can be found in the right diagram in figure \ref{fig:cont}. Let us discuss the effect of the new semi-circle on the two terms $\mathds{D}_a^{\Omega_0}$ and $\mathds{D}_b^{\Omega_0}$ arising from the expansion of the hyperbolic cotangent similar as in equation \eqref{eq:DtwotermsAp}:
\begin{itemize}
    \item The change of $\mathds{D}_b$ is now that the interval $\Omega \in [-\Omega_0,\Omega_0]$ is removed. This interval can be approximated in the following way (using that $\Omega_0 T_1 <\pi$, as discussed above):
    \begin{align}\label{eq:apbfdb}
        &\frac{C}{4\pi a^2}\frac{2a}{\pi}\text{Re} \sum_{n=1}^\infty \int_{-\Omega_0}^{\Omega_0} d\Omega \frac{a^2+\Omega^2}{a^2n^2 + \Omega^2} \frac{1}{\left(1-\frac{\Omega^2 T_1^2}{\pi^2}\right)^2} \left[ 1 - e^{i\Omega(\mathcal{T}+\mathcal{T}_1)} + e^{i\Omega \mathcal{T}_1} - \frac{1}{2} e^{i \Omega \mathcal{T}} -\frac{1}{2} e^{i \Omega (\mathcal{T}+2 \mathcal{T}_1)}\right] \nonumber\\
        &\leq \frac{C}{4\pi a^2}\frac{4\Omega_0 a}{\pi}\sum_{n=1}^\infty \frac{a^2+\Omega_0^2}{a^2n^2} \frac{4}{\left(1-\frac{\Omega_0^2 \mathcal{T}_1^2}{\pi^2}\right)^2} =\frac{2C }{3} \frac{\frac{\Omega_0}{a}\left( 1+ \frac{\Omega_0^2}{a^2}\right)}{\left(1-\frac{\Omega_0^2 \mathcal{T}_1^2}{\pi^2}\right)^2} \,,
    \end{align}
    where we used the absolute value and the triangle inequality. 
    This shows that for $\mathcal{T}\to\infty$, the contribution remains finite. As the new semi-circle forms a closed contour with the line from $-\Omega_0$ to $\Omega_0$ and does not contain any pole, its contribution must be equal to (the negative of) equation \eqref{eq:apbfdb} and therefore also remains finite for large $\mathcal{T}$.
    \item More interesting is the $\mathds{D}_a^{\Omega_0}$. To find out the effect of the cut here, we consider the case where $\Omega_0 \mathcal{T}_1 \ll 1$, 
    which can be achieved for suitable $\mathcal{T}_1$
    for all the cases we considered in the main text.
    The value of the semi-circle with radius $\Omega_0$ around the origin is
    \begin{align}
        &\text{Re} \frac{C}{4\pi^2 a} \int_0^\pi d\phi \; i \Omega_0 e^{i\phi} \frac{a^2+\Omega_0^2 e^{2i\phi}}{\Omega_0^2 e^{2i\phi}} \frac{1}{\left(1-\frac{\Omega_0^2 \mathcal{T}_1^2 e^{2i\phi}}{\pi^2} \right)^2}  \nonumber\\ &\hspace{1in}\cdot\left[ 1 - e^{i\Omega_0(\mathcal{T}+\mathcal{T}_1)e^{i\phi}} + e^{i\Omega_0 \mathcal{T}_1 e^{i\phi}} - \frac{1}{2} e^{i \Omega_0 \mathcal{T} e^{i\phi}} -\frac{1}{2} e^{i \Omega_0 (\mathcal{T}+2 \mathcal{T}_1) e^{i\phi}}\right]\nonumber\\
        &\approx  \text{Re} \frac{C}{4\pi^2 a\Omega_0} \int_0^\pi d\phi \; i \left(a^2 e^{-i\phi}+\Omega_0^2 e^{i\phi}\right) \nonumber\\ &\hspace{1.5in}\cdot \left[ 1 - e^{i\Omega_0(\mathcal{T}+\mathcal{T}_1)e^{i\phi}} + e^{i\Omega_0 \mathcal{T}_1 e^{i\phi}} - \frac{1}{2} e^{i \Omega_0 \mathcal{T} e^{i\phi}} -\frac{1}{2} e^{i \Omega_0 (\mathcal{T}+2\mathcal{T}_1) e^{i\phi}}\right]\,,
    \end{align}
    where we approximated $\left(1-\frac{\Omega_0^2 \mathcal{T}_1^2 e^{2i\phi}}{\pi^2}  \right)^2 \approx 1$. Now we have to integrate two different terms.
    \begin{itemize}
        \item The first term is of the form 
        \begin{align}
            \int_0^\pi d\phi \; e^{i\phi} e^{i b e^{i\phi}}= \int_1^{-1} dx \frac{-i}{x} x e^{i b x} =i \int_{-1}^1 dx e^{ibx} = \frac{2i}{b} \sin(b)\,.
        \end{align}
        We have substituted $x = e^{i\phi}$, 
        and we used that $b>0$ and that the transformation of the contour on the real line does not cross any pole (as this term arises from terms of the form $e^{i b x}$). In the limit $b\to 0$ it yields $\int_0^\pi d\phi\;  e^{i\phi}=2i$.
        \item The second term is of the form
        \begin{equation}
            \int_0^\pi d\phi \; e^{-i\phi} e^{i b e^{i\phi}}\,,
        \end{equation}
        which gives
        terms of the form $\frac{1}{x^2} e^{i b x}$. As this function has a singularity at the origin, we cannot transform the contour in the same way as above. Here, we close the contour by going on the real line to infinity, then following a semi-circle at infinite radius in the upper half plane and going from minus infinity to $-\Omega_0$. This contour does not include a pole and, also, the semi-circle vanishes, as 
        \begin{equation}
            \lim_{R\to\infty} \int_\pi^0 d\phi \frac{1}{R} e^{-i\phi}e^{i R b \cos(\phi)- R b \sin(\phi)} = 0\,.
        \end{equation}
        Hence it remains to evaluate the contribution of the function on the intervals $[-\infty,-\Omega_0]$ and $[\Omega_0,\infty]$ which is
        \begin{align}
            -i \int_{1}^{\infty} dx \frac{1}{x^2} e^{ibx} -i \int_{-\infty}^{-1} dx \frac{1}{x^2} e^{ibx} = -2i \int_1^\infty dx \frac{\cos(bx)}{x^2} = b\pi i - 2 i \cos(b) - 2 bi \text{SI}(b)\,,
        \end{align}
        where $\text{SI(x)}$ denotes the sine integral function. In the case of $b=0$, direct integration yields $\int_0^\pi d\phi \; e^{-i\phi} = -2i$. 
    \end{itemize}
    Collecting these results and inserting the corresponding values for $b$, we are left with
    \begin{align}\label{eq:newtermcutAp}
        \frac{C}{4 \pi^2 a} \Bigg[& -\frac{a^2}{\Omega_0} \Bigg( -2 + 2 \cos(\Omega_0(\mathcal{T}+\mathcal{T}_1)) -2\cos(\Omega_0 \mathcal{T}_1) + \cos(\Omega_0 \mathcal{T}) + \cos(\Omega_0(\mathcal{T}+2\mathcal{T}_1)) \nonumber\\ &\hspace{0.6in}-\Omega_0 (\mathcal{T}+\mathcal{T}_1) [\pi - 2\text{SI}(\Omega_0(\mathcal{T}+\mathcal{T}_1))] + \Omega_0 \mathcal{T}_1 [\pi - 2\text{SI}(\Omega_0 \mathcal{T}_1)] \nonumber\\ &\hspace{0.6in}-\frac{1}{2} \Omega_0 \mathcal{T} [\pi - 2\text{SI}(\Omega_0 \mathcal{T})] -\frac{1}{2} \Omega_0 (\mathcal{T}+2\mathcal{T}_1) [\pi -2\text{SI}(\Omega_0(\mathcal{T}+2\mathcal{T}_1))]\Bigg) \nonumber\\
        &-\Omega_0 \left( 2-2\frac{\sin(\Omega_0(\mathcal{T}+\mathcal{T}_1))}{\Omega_0 (\mathcal{T}+\mathcal{T}_1)} +2 \frac{\sin(\Omega_0 \mathcal{T}_1)}{\Omega_0 \mathcal{T}_1} -\frac{\sin(\Omega_0 \mathcal{T})}{\Omega_0 \mathcal{T}} - \frac{\sin(\Omega_0 (\mathcal{T}+2\mathcal{T}_1))}{\Omega_0 (\mathcal{T}+2\mathcal{T}_1)}\right) \Bigg]\,,
    \end{align}
    which is analyzed in specific limits in section \ref{sec:quhres2}.
\end{itemize}

\twocolumngrid

\bibliography{references.bib}

\end{document}